\documentclass{article}
\usepackage{graphicx}
\usepackage{xcolor}
\usepackage{parskip}
\usepackage{enumitem}%
\usepackage{comment}
\usepackage{caption}
\usepackage{float}
\usepackage{url}
\usepackage{booktabs}
\usepackage{color,soul}
\usepackage{tabularx}
\usepackage{amsmath}
\usepackage{bm}
\usepackage{url}
\usepackage{authblk}
\usepackage[numbers,sort&compress]{natbib}

\usepackage[a4paper, margin=1in]{geometry}

\usepackage{blindtext}

\begin{document}

\title{\huge\textbf{Accessible pore geometry governs tracer diffusion in crowded environments}}
\vspace{0.1in} 
\author[1]{Jinseok Lee}
\author[2]{Tong Lin} %
\author[2]{Mengyang Gu}
\author[1,*]{Yimin Luo}
\affil[1]{Department of Mechanical Engineering, Yale University, New Haven, CT 06511}
\affil[2]{Department of Statistics \& Applied Probability, UCSB, Santa Barbara, CA 93117}
\affil[*]{Corresponding author: yimin.luo@yale.edu}

\date{}
\maketitle

\begin{abstract}

Tracer diffusion in crowded environments is central to many biological and soft matter systems, but quantitative frameworks for linking tracer motion to environmental structure remain limited, and the co-dependence among geometric variables that facilitate or hinder tracer transport is not yet well understood. Here, we study the transport of rigid tracers in suspensions of soft particles and within living cells. Experiments reveal a transition from diffusive to confined motion as the matrix area fraction increases. The observed ensemble-level statistics, including the mean-squared displacements (MSDs), can be reproduced using a minimal simulation. Using simulation outputs, we train a parallel partial Gaussian process (PPGP) model that rapidly predicts MSDs from matrix geometric variables, including area fraction, particle size, and polydispersity. Analysis reveals that tracer transport is primarily governed by accessible pore sizes and that distinct global structures can produce indistinguishable MSDs.
While MSDs do not uniquely encode system geometrical parameters, we nevertheless find correspondence between pore size distribution and the ensemble MSDs. By modeling matrix self-diffusivity, the minimal model can also phenomenologically describe MSDs of internalized tracer particles in cells. 
The framework enables rapid inference of structural properties in crowded environments, including transport in the intracellular environment.

\end{abstract}

\section*{Introduction}

The transport of tracers amongst soft particles has vast biological implications \cite{lippincott2003photobleaching,reits2001fixed}. For instance, nutrient transport in intracellular environments can be hindered by macromolecules occupying up to 40 vol\% of the volume \cite{ellis2003join, zhou2008macromolecular,collins2019nonuniform}, and a similar degree of hindrance occurs in porous media \cite{ning2021diffusion, skaug2015hindered}. Tracer trajectories can provide both detailed spatial information and ensemble statistics. The former is useful for quantifying the final distribution of components during transport; the latter, in principle, can help reveal environmental properties using methods such as the generalized Stokes–Einstein relation using microrheology \cite{mason2000estimating,furst2017microrheology}. 

Particle trajectories have revealed unexpected phenomena that depart from classical Brownian descriptions where mean-squared displacement (MSD) grows linearly in time (Fickian), and the probability distribution of the displacement (i.e. the van Hove distribution) is Gaussian. Fickian yet non-Gaussian diffusion has been found in several biological and nonbiological settings \cite{wang2012brownian,wang2009anomalous,guan2014even,leptos2009dynamics,he2016dynamic} and attributed to spatial heterogeneities and time-dependent dynamics in the microenvironment \cite{guan2014even}. Later studies highlighted the key role of matrix mobility \cite{sentjabrskaja2016anomalous,sarfati2021enhanced}, linking non-Gaussian displacements to a transient short-time subdiffusive regime that precedes the long-time Fickian regime by several decades \cite{pastore2021rapid}. 
On the other hand, the van Hove can be Gaussian even when the MSD is nonlinear. This behavior is exemplified by fractional Brownian motion (FBM), in which a tracer's future motion is correlated with its past trajectory \cite{mandelbrot1968fractional,weiss2013single}. Although non-Markovian (not memoryless), FBM is typically ergodic, yet stochastic resetting can cause ergodicity breaking \cite{wang2021time}. FBM has successfully explained experimental observations of tracers viscoelastic media, such as protein \cite{banks2005anomalous} or actin solutions \cite{wong2004anomalous}, and within living cells \cite{weiss2004anomalous,tolic2004anomalous,golding2006physical,bronstein2009transient,sabri2020elucidating}.

Several other mechanisms are known to produce anomalous subdiffusive behaviors. 
In Continuous-Time Random Walks (CTRW) \cite{bouchaud1990anomalous}, the medium is homogeneous, but particles experience random waiting times; when the mean waiting time diverges, tracers become effectively trapped. Lorentz models instead describe particles moving through a forest of matrix particles. Near a critical matrix particle density, tracer motion becomes extremely slow. Incorporating polydispersity in this model impacts particle diffusion by changing the connectivity of the pore network, but the resulting diffusion can still be described by a universal percolation-based scaling law \cite{cho2012effect}. Unlike point-like tracers, finite-size tracers can become temporarily trapped at obstacle surfaces, leading to subdiffusive behavior \cite{zeitz2017active}. Hydrodynamic effects can be incorporated phenomenologically by replacing geometric exclusion with a position-dependent diffusion coefficient from Stokes flow, introducing lubrication-induced slowing near surfaces \cite{ermak1978brownian}. 
As the matrix area fraction $\Phi$ increases, a transition from diffusive to subdiffusive behavior is observed, whose nature depends on matrix particle polydispersity, size asymmetry between tracer and matrix particles, as well as the specifics of their interactions \cite{schnyder2017dynamic,schnyder2018crowding,cho2012effect,kurzidim2009single,skinner2013localization}. 

Despite these advancements, several fundamental questions remain unresolved. Tracer dynamics in crowded environments depend on multiple geometric variables: area fraction $\Phi$, matrix particle size $\bar{R}$, polydispersity $p$, and matrix particle mobility. 
Variation of these parameters in simulations has led to divergent conclusions about their impact on MSDs. For example, one study reports strong sensitivity of tracer diffusion to polydispersity at fixed particle size and area fraction \cite{cho2012effect}, while another finds that polydispersity has little effect on MSD scaling \cite{berry2014anomalous}.
Moreover, different physical mechanisms can produce nearly indistinguishable MSDs \cite{hofling2013anomalous}. Many crowded systems exhibit a generic sequence of dynamical regimes: free diffusion at short times, subdiffusion at intermediate times, and a crossover back to diffusion at long times. This behavior can arise from distinct physical origins, such as matrix mobility \cite{sarfati2021enhanced} or stochastic hopping dynamics \cite{xue2022hopping}, without explicitly resolving any geometric variables. Attempting to reconstruct features of a nonhomogeneous, structured environment from MSD remains challenging.

\begin{figure*}
\centering
\includegraphics[width=\textwidth]{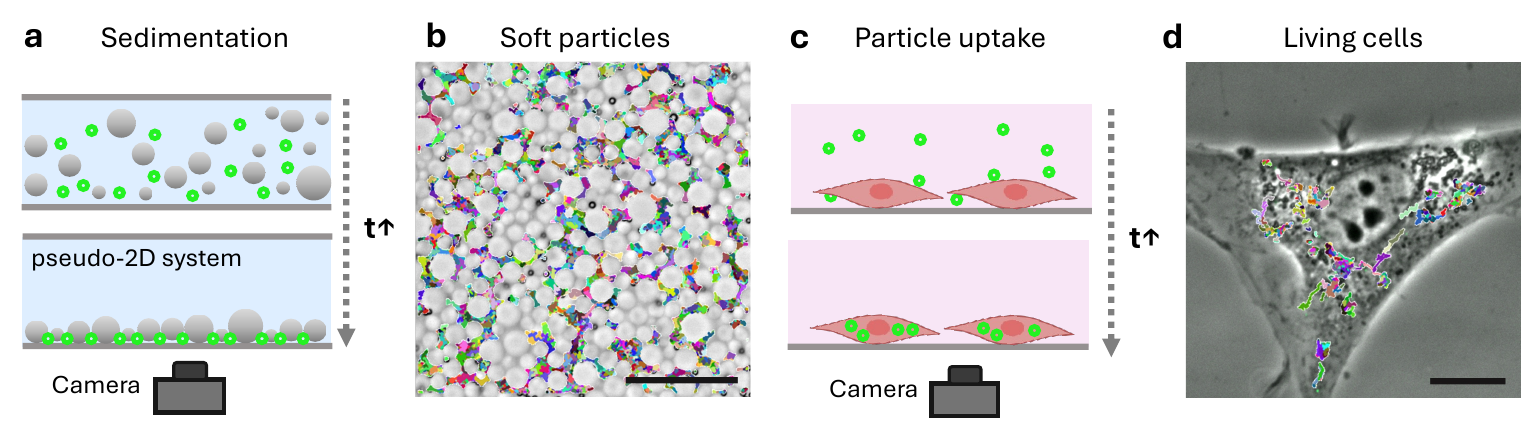}
   \caption{\textbf{Capturing ensemble tracer motility in passive and active crowded systems.} (a) A representative top-view image shows the soft matrix particles (gray) and hard tracers (green), overlaid with multicolored trajectories. (b) Matrix particle size distribution characterization. (c) Tracers internalized within living cells. (d) Trajectories are overlaid on a phase contrast micrograph of the cell. The scale bars are 50 $\mu$m in panel (b) and 20 $\mu$m in panel (d). }
\label{fig:experimental_setup}
\end{figure*}

To establish a more explicit link between system variables and observed dynamics, we first use a data-driven framework that incorporates $\Phi$, $\bar{R}$, and $p$ to construct a phenomenological model. The model enables the discovery of relationships between geometric variables and MSD. Then we ask which variables control the emergence of subdiffusive behavior and its crossover scales, and more broadly how much geometric information may be gleaned from tracer MSDs alone. Our experimental system consists of small, rigid tracers in both passive and active crowded environments (Fig. \ref{fig:experimental_setup}).  The passive system comprises tracers (2$a$ = 2 $\mu$m) dispersed among soft matrix particles with an average radius $\bar{R}$ = 4.35 $\mu$m in a pseudo-2D geometry (Fig. \ref{fig:experimental_setup}a,b). The active system consists of tracers (2$a$ = 100 or 200 nm) internalized by NIH3T3 cells (Fig. \ref{fig:experimental_setup}c,d).  
Using a parallel partial Gaussian process model \cite{gu2016parallel}, we predict thousands of MSD curves from input variables in seconds.
Rapid screening of the variable input space reveals that pore size distributions produce similar MSDs, whereas a sensitivity analysis confirms the interdependence of $\Phi$ and $\bar{R}$. 
Dynamic pore opening physically explains tracer jumps and escapes and accounts for the observed difference in MSD due to tracer size. 
We show that the model can phenomenologically describe transport in cells by accurately extracting realistic pore sizes, which we validate by reproducing the MSDs of tracers with different sizes.
Connecting system geometry to tracer transport will enable better formulation and understanding of transport in crowded environments. 

\section*{Results}
\subsection*{Tracer displacement in heterogeneous confinement: experiments and simulations}

\begin{figure*}[htb!]
\centering
\includegraphics[width=\textwidth]{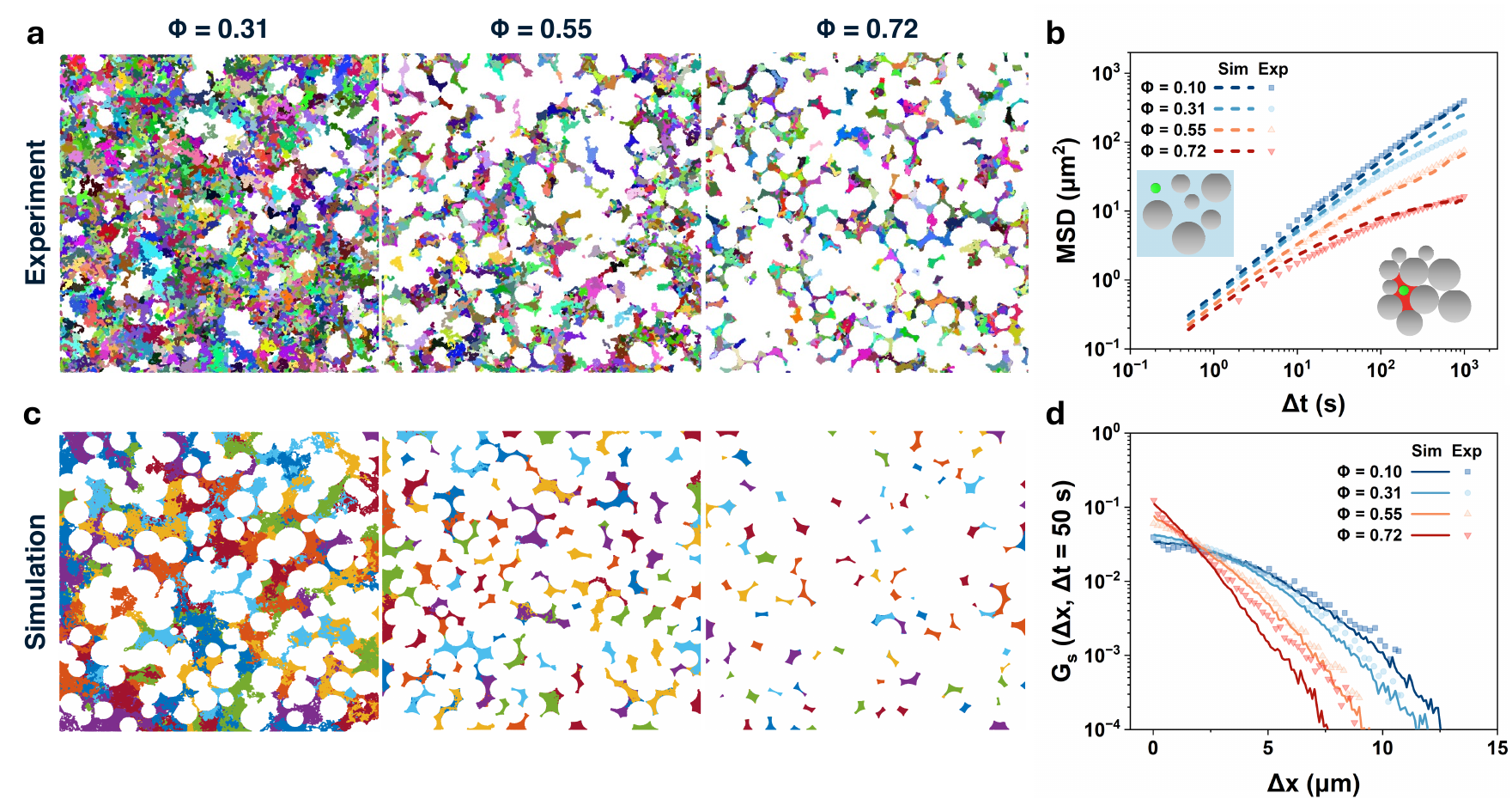}
   \caption{\textbf{Tracer motion becomes confined with increasing matrix area fraction.} (a) Representative tracer trajectories at increasing $\Phi$ for experimental samples prepared with 10 wt\% PEGDA. (b) Quantitative agreement between simulations and experimentally obtained MSDs. (c) Simulations using experimentally measured parameters, $\Phi$, $\bar{R}$ and $p$.
   (d) Self-part of the van Hove correlation function $G_s(\Delta x,\Delta t)$ at $\Delta t$ = 50 s, representative of the behavior observed at other $\Delta t$  values, following Eq. (\ref{eq:van_hove}) in Materials and Methods.} 
\label{fig:exp_diff_vol}
\end{figure*}

We track fluorescent polystyrene tracers ($2a = 2~\mu$m) dispersed between crosslinked PEGDA matrix particles, after they have sedimented to the bottom of a square glass capillary, forming a pseudo-2D layer (Fig.~\ref{fig:experimental_setup}a and \textbf{Supplementary Movie S1}). Given a density difference between water and polystyrene $\Delta \rho = 0.05 $ g/cm$^3$ is, thermal fluctuations would lift to a height of about $\Delta h =\frac{3k_BT}{4\pi a^3\Delta \rho  g} = 2\, \mu$m. Hence, the tracers are close to the bottom surface but remain suspended. Details on particle fabrication, imaging, and analysis procedures are provided in Materials and Methods. Representative trajectories of tracer particles at different matrix area fractions are shown in Fig.~\ref{fig:exp_diff_vol}a. As $\Phi$ increases, tracer motion becomes progressively confined, corresponding to a reduction in accessible pore space.

The experimental MSDs (shown as symbols in Fig.~\ref{fig:exp_diff_vol}b) are acquired by analyzing the tracer trajectories using both multiple particle tracking (MPT) and a differential dynamic microscopy (DDM)-based analysis called the ab initio uncertainty quantification (AIUQ) \cite{gu2024ab,lin2026model}. More details about MPT and AIUQ can be found in the Materials and Methods, and Section 1 of the supplementary text. We used AIUQ primarily as a supplementary check on the MSD obtained from MPT. In crowded matrices, tracer motion can exhibit intermittent hopping behavior, where particles remain confined within a pore for extended periods before transitioning to neighboring pores. This can complicate the choice of the search radius, an important parameter in MPT, and a different search radius is required in each scenario. 
Overall, we have verified that the MSDs agree using both methods (Fig. \ref{fig:exp_AIUQ_pred}, supplementary text).



Minimal models, such as the well-known Vicsek model \cite{vicsek1995novel}, are valuable tools for interpreting complex phenomenology. Owing to their simplicity, they can often be readily adapted and extended to account for phenomena in new settings \cite{chate2008modeling}. Rather than attempting to reproduce the full complexity of the experimental system, our goal is to identify the minimal set of physical ingredients required to reproduce the observed tracer dynamics. Here, we develop Monte Carlo simulations of $N_{t} = 200$ tracer particles moving among $N_m$ polydisperse matrix particles, for $T$ timesteps with a step size of $\Delta t$, in an $L\times L$ simulation box, scaled appropriately to match the experiment. The number of matrix particles $N_m$ is constrained by their area fraction $\Phi$, average radius $\bar{R}$, and polydispersity $p$. These variables were measured experimentally and used as inputs to the simulation (for details, see Section 2 of the supplementary text). The simulation makes it feasible to probe a broad range of the parameter space. 

To optimize the efficiency of the simulation, we implement several techniques. In the initialization step, we populate the box with matrix particles in descending size until the target $\Phi$ is reached (\textbf{Supplementary Movie S2}). This procedure is analogous to filling a jar first with large rocks, then with gravel, and eventually with sand. If an insertion fails before reaching $\Phi$, a local relaxation is applied: existing particles first undergo a Brownian random walk. Then their positions are updated via a pairwise repulsive potential, and finally any remaining overlapping pairs are displaced along their line of centers (Section 2.1 of the supplementary text). Once an initial matrix particle configuration is reached for specified $\Phi$, $\bar{R}$, $p$, tracer particles are allowed to move in the interstitial sites between the matrix particles through position updates. Whenever a tracer particle crosses the boundary of a matrix particle, the step is resampled until no overlaps are detected.

If the tracer moves freely in the medium (water), its diffusion coefficient follows $D_0 = \frac{k_BT}{6\pi\eta a}$, where $\eta$ is the viscosity of the medium. However, tracers are close to the bottom surface and to matrix particles, and these confinements reduce their mobility.
The hindrance from the glass substrate is obtained by experimentally measuring the perpendicular diffusivity $D_{\perp}$ without any matrix particles, which is smaller than $D_0$. This substrate-induced correction applies regardless of the tracer position relative to the matrix particles.
A second contribution comes from the tracer’s proximity to matrix particle surfaces, which induces lubrication-like hydrodynamic drag and further suppresses tracer mobility \cite{KimKarrila1991}. We model this effect through a gap-dependent correction, producing a final position-dependent effective diffusivity,
$D^*\sim{D_\perp}\left(\frac{g^\ast}{g^\ast + g_0}\right)^{\alpha_{\mathrm{hyd}}}$,
where $g^\ast$ is the surface-to-surface gap between the tracer and the nearest matrix particle, and $g_0$ is a characteristic length scale that sets the onset of hydrodynamic hindrance. Unlike the geometric variables $\Phi$, $\bar{R}$, and $p$, the hydrodynamic parameters $\alpha_{\mathrm{hyd}}$ and $g_0$ are held constant across all simulation conditions. This treatment is phenomenological: it is intended to consistently capture the general effect of tracer slowdown near matrix particles, rather than providing a complete hydrodynamic description. Full details of the model are provided in Section 2.2 of the supplementary text, and a representative video is shown in \textbf{Supplementary Movie S3}.

Simulated particle trajectories are shown in Fig.~\ref{fig:exp_diff_vol}c, which capture the gradually decreasing accessible space for the tracers for increasing $\Phi$. Simulated MSDs are shown as dashed lines in Fig.~\ref{fig:exp_diff_vol}b.  We find that the experimentally observed MSD at different $\Phi$'s can generally be reproduced using the model with fixed parameters (Table \ref{tab:sim_params}). The simulations also recapitulate the experimentally measured self-part of the van Hove correlation function (Fig.~\ref{fig:exp_diff_vol}d). In addition, we also plotted the non-Gaussian parameter $\alpha_2$ and the velocity autocorrelation function (VACF), defined in Section 3 of the Supplementary Text, and shown in Fig. \ref{fig:non_Gaussian_param_VACF}. 
The agreement between experiment and simulation demonstrates that the model reproduces the full distribution of tracer displacements, rather than only the MSDs. Over the observed timescale, the processes appear to be both non-Gaussian and non-ergodic, and they share similarities with a CTRW process, in which the tracer spends a long time in pores between successive jumps. A similar interpretation was reported in Ref. \cite{regner2013anomalous}. Nonetheless, CTRW treats jumps through a mean-field approximation \cite{bouchaud1990anomalous} and does not explicitly account for the underlying geometry. In contrast, our simulation can predict MSD behavior directly from geometric variables. 

\subsection*{Accurate prediction of simulation by data-driven predictive models}

\begin{figure*}
\centering
\includegraphics[width=\textwidth]{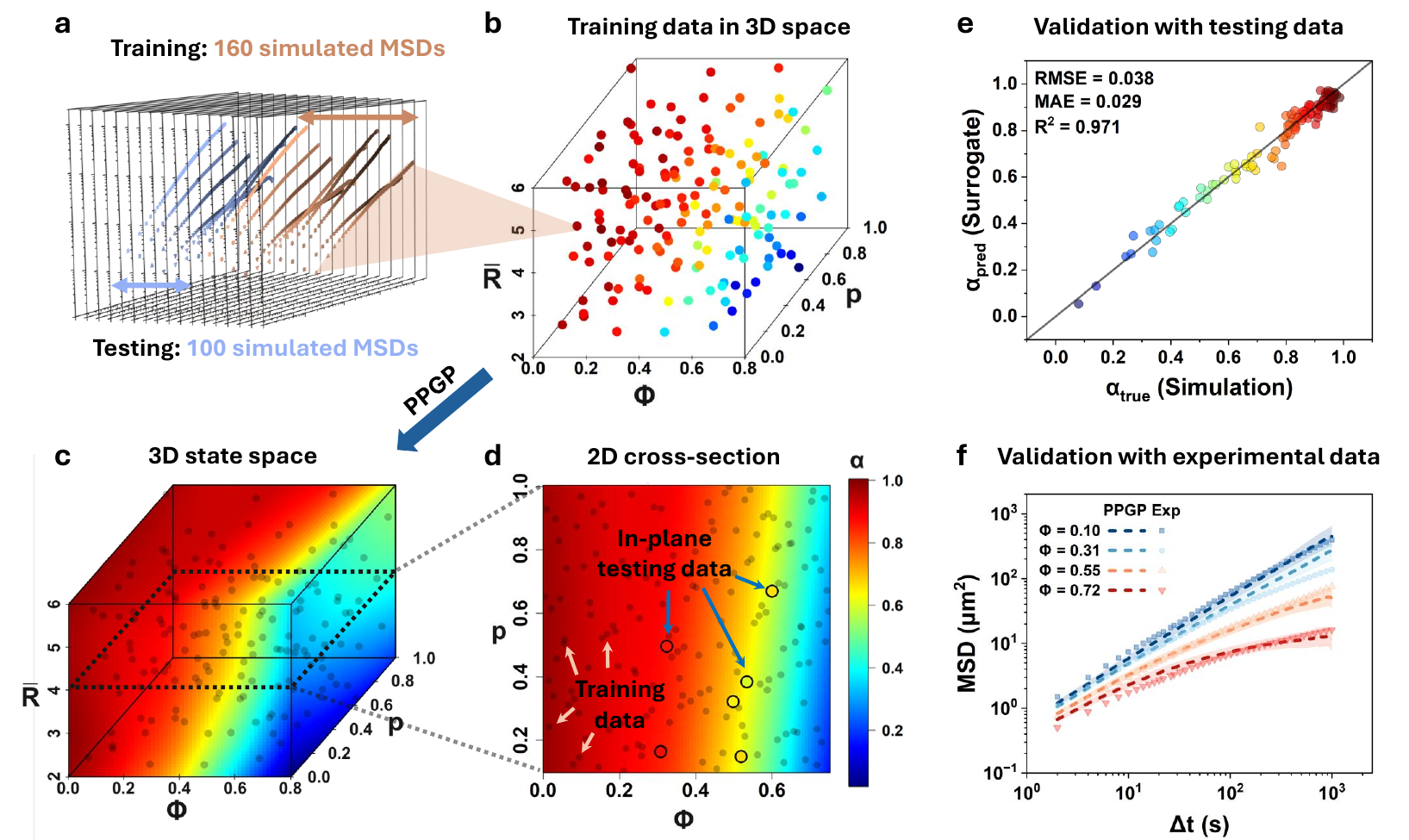}
   \caption{\textbf{Illustration of the PPGP predictive model pipeline.} (a) The simulations use geometric inputs ($\Phi$, $\bar{R}$, $p$) to compute the MSDs. (b) The MSDs are visualized by one-dimensional $\alpha$ by computing the least-squares slope of  $\log\langle \Delta r^2\rangle$ versus $\log\Delta t$ at different locations. (c) Reconstructed 3D state space in which each point represents a 500-dimensional MSD (gray points denote training data). (d) 2D cross-section where the testing data (outlined in black) are similarly color-coded, showing agreement with the PPGP predictive model. (e) Validation with testing data. Model performance is quantified using RMSE, MAE, and $R^2$ (see definitions in Section 4 of the Supplementary Text). (f) The PPGP model, trained exclusively on simulation data, is validated using experimental observations. The shaded areas denote the 95\% interval computed from the PPGP predictive model.}
\label{fig:PPGP_pipeline}
\end{figure*}

To understand relationships between MSD curves and model inputs in the entire input space, we build 
a parallel partial Gaussian process (PPGP) model \cite{gu2016parallel}, a data-driven predictive model to predict the vector of MSD curves generated from the simulations with inputs $\Phi$, $\bar{R}$, and $p$. 
Unlike conventional Gaussian process regression \cite{rasmussen2006gaussian}, which predicts functions with scalar-valued outputs, the PPGP  model is designed to predict functions with vectorized outputs based on a small number of training data. 
In the PPGP model, the MSD curve at each lag time follows a Gaussian process with distinct mean and variance parameters estimated from the data, which is crucial for this scenario, as the scales of the MSD curve at different lag times can differ.  Furthermore, the correlation function between each output value of the MSD curves over distinct geometric inputs ($\Phi$, $\bar{R}$, $p$) is assumed to be the same in the PPGP model, which reduces the computational complexity and increases the stability of the estimation.   
We use the RobustGaSP package \cite{gu2018robustgasp,gu2018jointly} to estimate the parameters in the PPGP model based on 160 simulated MSD curves (Fig.~\ref{fig:PPGP_pipeline}a), and use predictive mean and quantiles of MSD curves from the PPGP model for prediction and uncertainty quantification, respectively \cite{gu2016parallel}. 
For visualization purposes only, the MSDs are represented by a single quantity $\alpha$ (the logarithmic slope of the MSD, Fig.~\ref{fig:PPGP_pipeline}b), when in reality it is a function of $\Delta t$. A single colored dot represents the full 500-dimensional output from the prediction model, and is shown only to visualize 2D outputs in the 3D input parameter space.

The PPGP  model predicts MSD curves across the three-dimensional input space defined by $\Phi$, $\bar{R}$, and $p$, with the predictive logarithmic slope plotted in Fig.~\ref{fig:PPGP_pipeline}c. The predictions in a 2D-projection plane are compared with six in-plane held-out test data (Fig.~\ref{fig:PPGP_pipeline}d, color-coded and outlined in black), which show good agreement with the prediction from the PPGP predictive model. We further validate the model by comparing the predicted $\alpha$ values with the slopes of MSD obtained directly from simulations for all 100 held-out test samples (Fig.~\ref{fig:PPGP_pipeline}e), with a correlation coefficient $R^2$  = 0.971. 
We verify that the PPGP predictive model, trained only with inputs from the simulation data, can accurately predict the experimental MSD (Fig.~\ref{fig:PPGP_pipeline}f). It generates an MSD using the experimentally measured variables $(\Phi, \bar{R}, p)$ in less than 25 ms on a laptop.
Constructing a state space allows us to explore a larger input space that would be too cumbersome to investigate experimentally.  

\begin{figure*}
\centering
\includegraphics[width=0.75\textwidth]{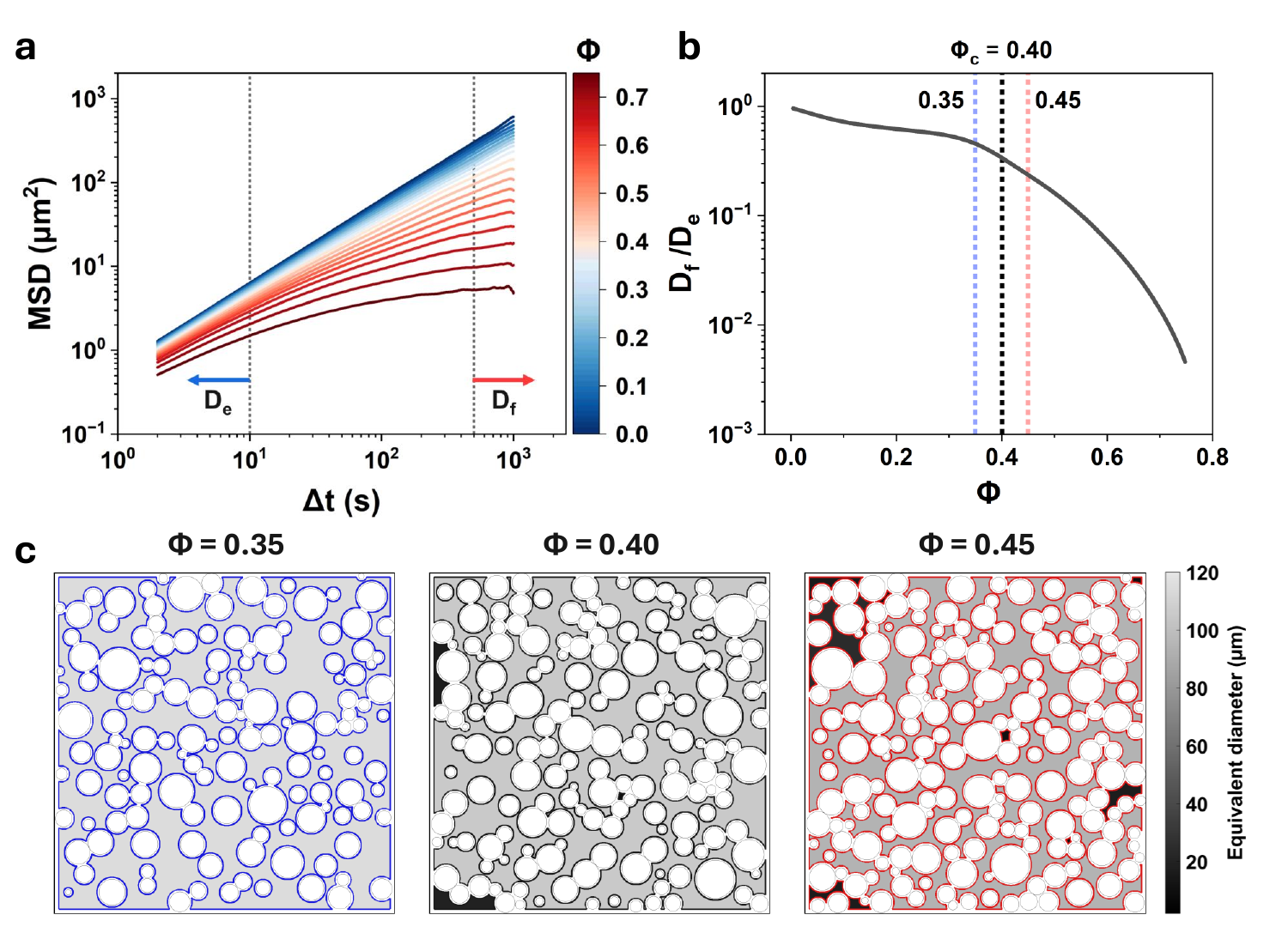}
   \caption{\textbf{The PPGP predictive model enables rapid identification of the tracer-defined percolation threshold.} (a) MSDs predicted by the PPGP model for different area fractions $\Phi$ at fixed values of $\bar{R}$ and $p$. The early-time and late-time diffusion coefficients are denoted as ${D_\mathrm{e}}$ and ${D_\mathrm{f}}$, respectively. Dashed lines indicate the time intervals $\Delta t$ over which a linear regression is performed to fit the MSD. (b) Determination of the critical area fraction $\Phi_c$, defined as the ``knee'' of the curve $\frac{D_\mathrm{f}}{D_\mathrm{e}}$ versus $\Phi$.
   (c) Snapshots of the system immediately before and after $\Phi_c$. Below $\Phi_c$, no trapped pores are present; above $\Phi_c$, trapped pores start to appear and grow.}
\label{fig:superposition}
\end{figure*}

Because data-driven prediction of simulation outputs is rapid and accurate, we can identify the percolation threshold $\Phi_c$ for a specific mixture with fixed $\bar{R}$ and $p$. Here, we generate 300 MSDs (20 are shown in Fig.~\ref{fig:superposition}a for clarity) for a different matrix size distribution
($\bar{R}$ = 3.76, $p$ = 0.500), which only takes $\sim$7.5 seconds by the PPGP predictive model. Previous studies \cite{bi2016motility} have relied on the logarithmic slope of the terminal MSD to define the transition. However, this approach may overlook the initial confinement present in the system, which we find may vary slightly depending on system geometry. Our procedure accounts for this, as described in detail in Section 5 of the Supplementary Text. Briefly, we first estimate the diffusion coefficients $D_\mathrm{e}$ and $D_\mathrm{f}$ for both small and large $\Delta t$'s. Then we compute their ratio, which is a function of area fraction $\Phi$. The ratio can be interpreted as the reduction in displacement due to confinement, where $\frac{D_\mathrm{f}}{D_\mathrm{e}}$ = 1 in the absence of confinement.  We then apply the kneedle algorithm to identify the knee of this curve, $\Phi_c$, which represents the point of maximum curvature (Fig.~\ref{fig:superposition}b). We note that although the $\Phi_c$ identified here is related to previously reported jamming transitions \cite{o2003jamming}, it is not identical to the jamming transition of the matrix itself. Instead, it is defined from the tracer’s perspective and therefore explicitly incorporates the tracer’s excluded volume. As a result, regions of inaccessibility emerge at lower area fractions, since the tracer can be excluded from narrow necks between matrix particles. Consequently, the tracer-defined threshold is expected to occur prior to the geometric percolation transition of the void space or the jamming transition of the matrix itself.  At $\Phi_c$, we begin to observe trapped pore space, which grows with increasing $\Phi$ (Fig.~\ref{fig:superposition}c). This tracer-defined point provides a distinct and practical way to find the area fraction when tracers become trapped. 

\subsection*{The tracer displacement is governed by the available pore space}

\begin{figure*}
\centering
\includegraphics[width=\textwidth]{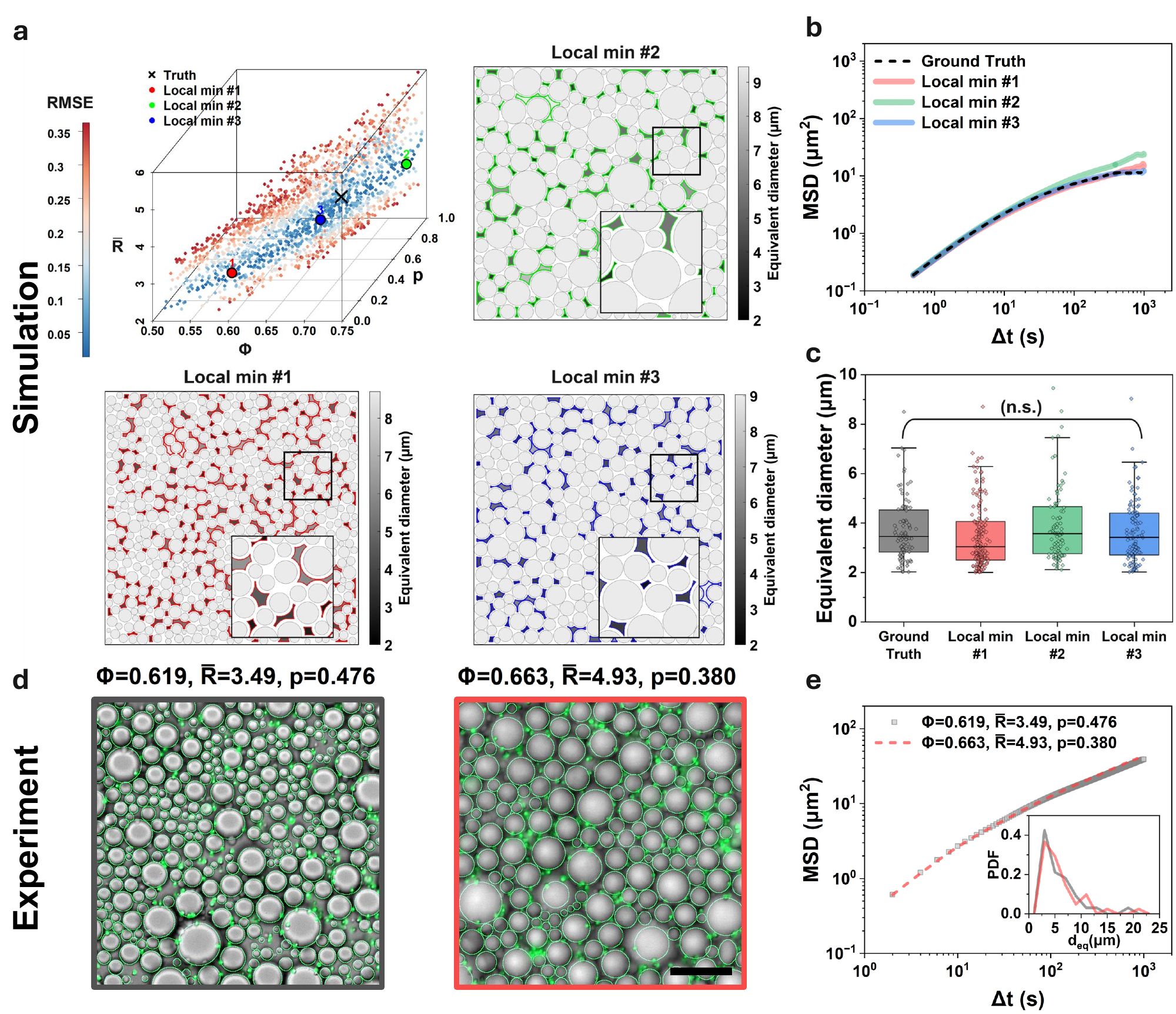}
   \caption{\textbf{Distinct matrix geometries can produce nearly indistinguishable tracer MSDs.} (a) The root mean square error (RMSE) of the sampled MSD compared to the ground truth MSD (marked by $\times$) located in the parameter space. Three possible parameter configurations (colored circles) are highlighted and shown as insets. Local min \#1: $\Phi$ = 0.576, $\bar{R}$ = 2.58, $p$ = 0.260, Local min \#2: $\Phi$ = 0.728, $\bar{R}$ = 3.57, $p$ = 0.958, Local min \#3: $\Phi$ = 0.698, $\bar{R}$ = 4.14, $p$ = 0.214. (b) The MSDs corresponding to each of these three conditions are plotted as \textit{symbols}, compared to the ground truth MSD, which is shown as the dashed line. (c) A box plot shows the distribution of pore areas.
  (d) Two experimental conditions with different $\Phi$, $\bar{R}$, and $p$. The scale bar is 30 $\mu$m (e) Comparable MSDs are observed in both cases. The inset shows the probability density function of the pore equivalent diameter ($d_{eq}$).}
\label{fig:inverse_prob}
\end{figure*}

In principle, the geometric variables ($\Phi$, $\bar{R}$, $p$) can be measured from the microscopy videos, but in practice, filtering and isolating features from the background cannot be done with high accuracy automatically, and tallying a polydisperse system manually can take up to an hour per dataset. Therefore, this provides a basis for identifying plausible structures by using the PPGP predictive model to map the MSDs back to nonunique sets of geometric variables. We evaluate this approach using a simulated MSD (ground truth) that was not included in the training dataset. This target is represented by a ``$\times$'' symbol in the 3D parameter space. We sample 12,000 MSDs throughout the 3D space ($\sim$5 minutes) and compute the root mean square error (RMSE) between the target MSD $\langle \Delta r^2\rangle$ and each sampled MSD $\langle \Delta \hat{r}^2\rangle$($\Phi$, $\bar{R}$, $p$):
\begin{equation}
\mathrm{RMSE}
=
\sqrt{
\frac{1}{500}
\sum_{i=1}^{500}
\left(
\langle \Delta r^2(\Delta t_i) \rangle
-
\langle \Delta \hat{r}^2(\Delta t_i) \rangle
\right)^2
}
\end{equation}
The RMSE is color-coded, where blue regions correspond to lower RMSE and closer agreement with the target MSD. For clarity, only data points with RMSE $\leq 0.35$ are shown (Fig.~\ref{fig:inverse_prob}a).
This analysis shows that many points in the parameter space produce nearly identical MSDs. The lowest RMSE region in the parameter space appears as a continuous band. We visualize three representative points amongst them, each with distinct particle distributions (Fig.~\ref{fig:inverse_prob}a, insets). Their corresponding MSDs are plotted in Fig.~\ref{fig:inverse_prob}b, showing that they all resemble the ground truth. 
By analyzing the pore size (details can be found in Section 6 of the supplementary text), we find that configurations \#1 to \#3 yield pore size distributions that are virtually indistinguishable, despite corresponding to distinct global configurations (Fig.~\ref{fig:inverse_prob}c). This implies a correspondence between pore size distribution and MSD. The reason is that each particle probes only its immediate surroundings, so its trajectory reflects local geometric features rather than the global geometry. This conclusion, generated entirely using simulated data, is validated experimentally: two distinct matrix particle distributions (Fig. \ref{fig:inverse_prob}d) with the same pore size distribution (Fig. \ref{fig:inverse_prob}e, inset) produce identical MSDs (Fig. \ref{fig:inverse_prob}e).

To understand which variables most strongly drive the changes in MSD, we first standardize each variable by computing the $z$-statistics: $X_\mathrm{std} = \frac{X-\mu_X}{\sigma_X}$, where $X$ denotes either the output variable RMSE or one of the input variables $\Phi$, $\bar{R}$ or $p$, $\mu_X$ stands for the mean, and $\sigma_X$ is the standard deviation of $X$. The standardized variables are fit to a multivariate regression:
\begin{equation}
    \mathrm{RMSE}_\mathrm{std} = \beta_{\Phi}\Phi_\mathrm{std}+\beta_{p}p_\mathrm{std}+\beta_{\bar{R}}\bar{R}_\mathrm{std}+\epsilon,
\end{equation} 
where $\epsilon$ denotes Gaussian noise with a mean of zero. 
The variable with the largest standardized coefficient $\beta$ is expected to have the strongest influence on RMSE. The fitted coefficients are $|\beta_{\Phi}|=0.8654$, $|\beta_{\bar{R}}|=0.1902$, and $|\beta_{p}|=0.0907$ ($R^2=0.7978$). Among the three variables, $\Phi$ emerges as the dominant factor, accounting for most of the observed variability, whereas $p$ plays a comparatively minor role.

To understand nonlinear effects and interactions between inputs, we also perform a global sensitivity analysis using Sobol’s indices \cite{sobol1990sensitivity}, a variance-based measure that quantifies each variable's individual and joint contributions to the variance of RMSE. This is done by using the sensitivity package in R \cite{pujol2017sensitivity}. This analysis produces first-order sensitivity indices $S_p$ = 0.051, $S_{\Phi}$ =0.086, and $S_{\bar{R}}$ = 0.172, and second-order sensitivity indices $S_{p\Phi}$ = 0.014, $S_{\Phi\bar{R}}$ = 0.578, and $S_{\bar{R}p}$ = 0.038,  
which reveals that the interaction between $\bar{R}$ and $\Phi$ has the greatest effect on the accuracy of the MSD. 
These results indicate that $\bar{R}$ also plays an important role, because it influences pore size and is correlated with $\Phi$. 

\subsection*{Modeling transport in the intracellular environment}

The cell cytoplasm is a densely crowded environment: beyond macromolecules \cite{ellis2003join}, it is further structured by the actin cortex, which has a mesh size of 100-200 nm \cite{svitkina2020actin}, and by microtubule networks, which have a mesh size on the order of 1 $\mu$m \cite{burakov2021persistent}. 
Nanoparticles (NPs), increasingly employed as drug delivery carriers, typically enter the cytoplasm via endocytosis and must navigate this crowded intracellular space. Their transport is affected by binding interactions with macromolecules, active transport processes driven by molecular motors, spatial heterogeneity due to placement of the organelles, and viscoelasticity of the cytoskeleton.
Notably, previous work has analyzed tracer trajectories to infer cellular mechanical properties by tracking the thermal and active transport of probes and macromolecules, also known as biomicrorheology \cite{wirtz2009particle,weihs2006bio,hale2009resolving}. These experiments concluded that observed tracer speeds were too low to be mediated by molecular motors. Still, early interpretations lacked a unified framework to explain the resulting dynamics \cite{weihs2006bio}. More recent work helps bridge this gap by establishing that the cytoplasm behaves as a poroelastic material, a porous elastic network permeated by cytosolic fluid \cite{moeendarbary2013cytoplasm}. Within this framework, crowding and geometric constraints on the accessible free volume are sufficient to reproduce key features of macromolecular transport \cite{destrian2026cytoplasmic}. Despite these advances, a quantitative connection between the observed single-particle trajectories and local geometry remains incomplete.

\begin{figure*}
\centering
\includegraphics[width=\textwidth]{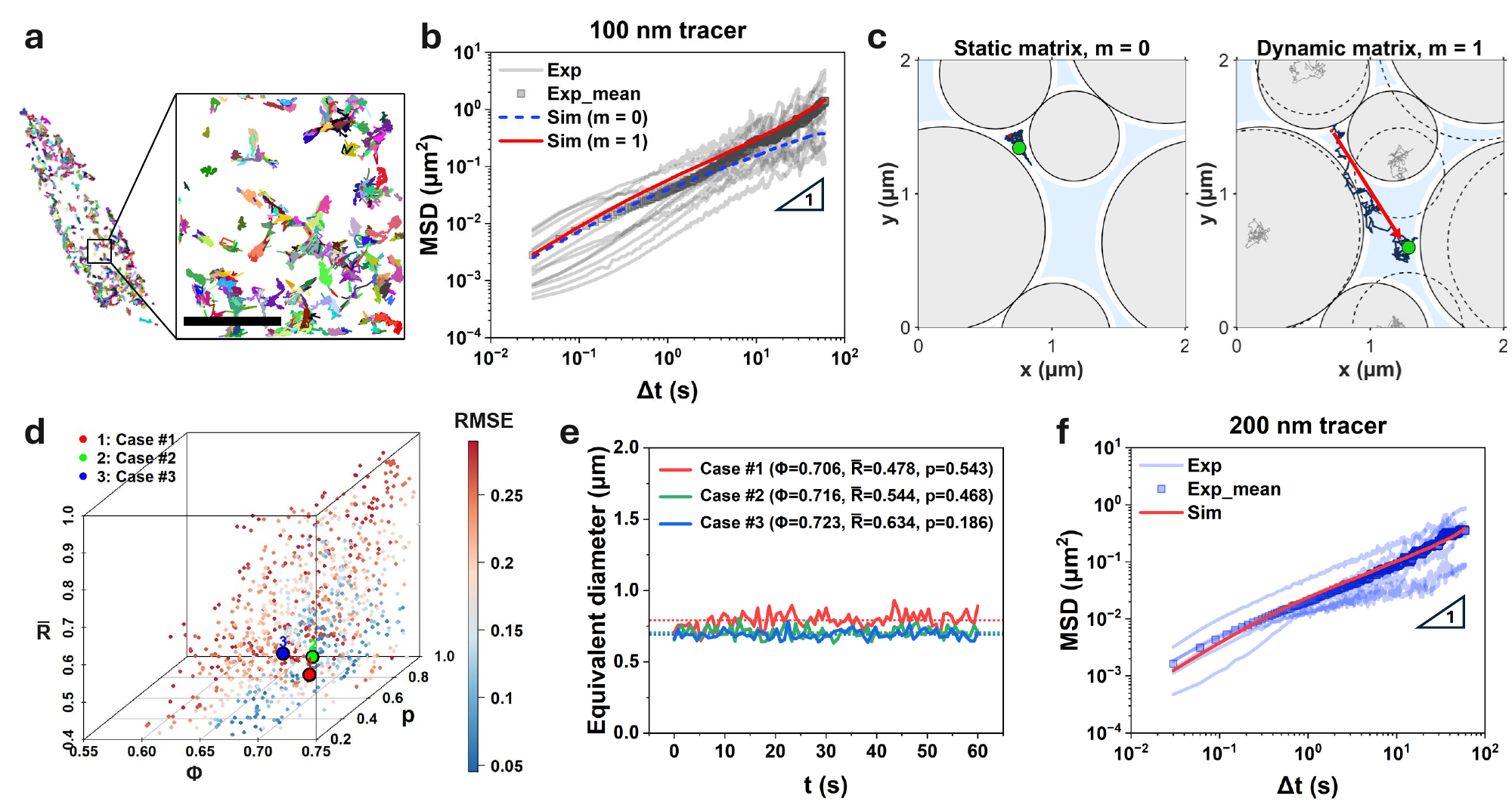}
   \caption{(a) Representative trajectories of intracellular tracers within a single NIH3T3 fibroblast cell. The inset shows a magnified view of the highlighted region. (b) MSDs averaged over 40 tracers (2$a$ = 100 nm) in 13 different cells. Cell-averaged MSDs are shown in gray. (c) Example tracer trajectories (thick navy lines) from static versus mobile pores. Thin gray lines show matrix particle trajectories, and dashed lines denote their final positions. The time interval between the initial and final configurations is $\Delta t = 3~\mathrm{s}$. (d) RMSEs by fitting the experimental tracer MSD.  The three cases with the lowest RMSE values are highlighted. (e) All three cases exhibit similar pore sizes. (f) Validation using a different tracer size (2$a$ = 200 nm, averaged over 6 different cells) demonstrates agreement between simulations based on the extrapolated variables and experimental measurements. The scale bars are 10 $\mu$m in panel (a) and 4 $\mu$m in panel (b).}
\label{fig:cell_surrogate}
\end{figure*}

Here, we ask whether a minimal description that accounts only for the available pore space can capture the dominant trend in tracer motion.
This is intended to be a phenomenological demonstration that effective pore-scale geometric constraints can reproduce aspects of the observed transport behavior, specifically, how ``accessible'' different regions of the cytoplasm are to diffusing tracers, rather than proof that porosity alone governs intracellular diffusion. We use carboxylated tracers (2$a$ = 100, 200 nm), which are naturally internalized by NIH3T3 cells through endocytosis, as the model system. More information on sample preparation is presented in the Materials and Methods. We find that tracers tend to avoid certain regions (Fig.~\ref{fig:cell_surrogate}a), consistent with transport in heterogeneous media containing immobile obstacles.
We average a large number of trajectories to capture the average tracer behavior in the heterogeneous cytoplasm. 

Throughout the $\Delta t$ range probed, MSD is sub-diffusive, indicative of hindered motion consistent with transient interactions with the actin cortex and other intercellular components (Fig.~\ref{fig:cell_surrogate}b).  The tracer trajectories show no persistent drift in any direction and are instead dominated by stochastic fluctuations throughout the observation window. No significant morphological changes in cell shapes are detected either, suggesting that large-scale active rearrangements do not dominate the measured statistics. 
The observed trajectories are consistent with confinement in local pore-like regions,  followed by escape and diffusion at long lag time. The confinement is transient, a feature not taken into account in the static pore model (blue dashed line, Fig.~\ref{fig:cell_surrogate}b). To simplify the modeling, we represent the dynamic rearrangement of the pore spaces through the self-diffusion of the matrix particles, similar to Refs. \cite{sarfati2021enhanced,sentjabrskaja2016anomalous,berry2014anomalous}.

Reported values of cytoplasmic viscosity vary widely from 1-$10^6$ times that of water \cite{kuimova2009imaging,betterton2022new,berret2016local}. Here, the best fit to the data is when $\eta\approx$ 100 times that of water, which lies within the aforementioned range and serves as an order-of-magnitude estimate. Note that this effective viscosity is a coarse-grained parameter for general dissipation, 
that will produce Brownian-like motion through combined effects of cytoplasmic viscosity, viscoelasticity, and active forces, rather than thermal flucutations. We also assign each matrix a diffusivity $D_M = m\frac{k_BT}{6\pi\eta R}$, where $R$ is the matrix particle radius and $m$ is the mobility factor. Because the matrix particles are confined by their neighbors, their actual movement is substantially restricted. The effect of varying matrix mobility is shown in Fig. \ref{fig:effect_of_mobility}, which illustrates that different $m$ does not significantly change the tracer MSD. For simplicity, $m$ is set to 1 in the subsequent discussions.
Allowing matrix particles to rearrange captures certain transport behavior of a poroelastic network, including dynamic evolution of pore geometry and accessible space. A representative tracer trajectory is shown in Fig. \ref{fig:cell_surrogate}c, illustrating confinement within a static pore (left) and escape through a dynamically opening pore (right). 


Because reconstructing the input variables from the MSD does not yield a unique solution, we instead try to identify variable sets ($\bar{R}$, $\Phi$, $p$) that plausibly reproduce the observed MSDs using PPGP (Fig. \ref{fig:cell_surrogate}d), as before. In this case, low- and high-RMSE outputs are more interspersed despite retaining some correlation amongst the variables. This is attributed to dynamic connectivity, which exposes tracers to varying instantaneous accessible space, thereby increasing the sensitivity of the MSD to local fluctuations. As an illustration, we select 3 representative local minima from this parameter space and show that they have similar time-averaged pore sizes (Fig. \ref{fig:cell_surrogate}e). The extracted pore size is comparable with the reported $\sim 1~\mu\mathrm{m}$ mesh size derived from fluorescent microscopy and DDM analysis \cite{saldanha2025competing,burakov2021persistent}. This value reflects the effective accessible void space experienced by tracers rather than directly representing organelle spacing. These observations indicate that similar MSDs can arise from similar pore structures even when the pore dynamically rearranges.

If the tracer motion is indeed governed by the local geometry characterized by pores, then the same set of variables derived from the MSD analysis should also reproduce MSDs under a new set of experimental conditions. Since we cannot easily change the composition of the cytoplasmic makeup, we instead increase the tracer size to 200 nm, to test the predictive capability of the model through extrapolation in tracer size. We rerun the simulation by doubling the tracer size, while keeping all other parameters fixed. The resulting MSD matches the experimentally measured, ensemble-averaged MSDs of larger endocytosed tracers (2$a$ = 200 nm, Fig.~\ref{fig:cell_surrogate}f). While the unbounded diffusion coefficient is expected to halve with increasing tracer size, the simulation also captures the characteristic curvature changes in the MSD induced by near-field effects. As a result, the trajectories of the 200 nm tracers are not merely a shifted version of those from the 100 nm tracers. At the longest $\Delta t$, 100 nm particles are diffusive, whereas 200 nm particles remain subdiffusive. Indeed, interactions with membranes, cytoskeletal structures, and active transport machinery could also contribute to the observed trajectories. However, the goal of the cellular simulations is not parameter inference or quantitative reconstruction of intracellular physics, but rather to test whether biologically reasonable parameter regimes can reproduce qualitative features of the measured tracer dynamics. 


\section*{Discussion}



Studying transport in crowded soft environments is important for understanding processes ranging from intracellular trafficking to diffusion through gels, emulsions, and porous materials. Understanding such interactions enables one to predict when transport will be unhindered, slowed, or completely jammed, which is critical for designing better drug delivery systems, improving the manufacturing of gels and emulsions, and devising strategies to enhance transport in crowded cellular environments, as most organelles can be modeled as soft, polydisperse, self-diffusing particles. 

In this work, we model the transport of tracers within a matrix of soft, polydisperse particles through experiments and modeling. Increasing matrix area fraction drives tracer trajectories from diffusive to subdiffusive motion, indicative of caging effects. A minimal simulation captures the experimentally observed MSD using measured system geometric variables. 
We then use a data-driven predictive model to map geometric variables onto MSD, enabling efficient phase space exploration. 

We find that multiple regions of parameter space can produce nearly identical MSDs, and the inverse inference from MSD alone is not unique. Such degeneracy arises due to similar pore size distributions. This physical insight is gained by a model trained entirely on simulation data, but is also confirmed by experiment. A global sensitivity analysis reveals a correlation between the input variables. The observed degeneracy between matrix parameters and tracer dynamics provides data-driven support for the earlier conclusion that MSD alone is insufficient to uniquely determine the underlying physical mechanisms \cite{hofling2013anomalous,ernst2014probing}. Rather than establishing a one-to-one inverse mapping, the present work identifies geometrically distinct but transport-equivalent environments. Notably, both the experimental systems generated by sedimentation and the simulated matrices generated via the jar-filling procedure exhibit similarly disordered structures, which likely contributes to the agreement between experiment and simulation. As a result, this work does not establish that ($R$, $\Phi$, $p$) are sufficient descriptors for more structured porous environments, where transport may additionally depend on higher-order spatial correlations, such as ordered lattices or anisotropic matrices. It instead shows that accessible pore-scale provides a physical basis for intracellular tracer dynamics in randomly packed, polydisperse systems.

Beyond macromolecular crowding, the intracellular environment is highly complex and also includes active remodeling, viscoelasticity, molecular binding, and nonequilibrium fluctuations, none of which are explicitly modeled in the present framework. Multiple mechanisms likely coexist in cells \cite{hofling2013anomalous}. Notably, viscoelasticity can also generate intermediate-time MSD plateaus similar to those observed experimentally, and the present study does not uniquely distinguish between these mechanisms. Nevertheless, here we investigate whether effective crowding and pore-scale constraints alone can account for some aspects of the observed MSD behavior of endocytosed tracers, by capturing the dominant features of their dynamics. We find that matrix particle motility is necessary to recapitulate the long-time MSD. This motion alters the connectivity of the pore network and increases the MSD’s sensitivity to local fluctuations. Nevertheless, we find that the correspondence between pore size distribution and MSD holds, even when the matrix is dynamic and pore connectivity changes. Using MSDs measured with one tracer diameter ($2a$ = 100 $\mu$m), we infer the distribution of pore sizes, and validate the results using a second tracer size ($2a$ = 200 $\mu$m). Then we rerun the simulation and validate against particle tracking MSD for this larger diameter, without modifying any other simulation parameters. The agreement of their MSDs suggests that, although the system is highly stochastic, the inferred structural properties of the medium are robust and can reliably predict tracer dynamics across different particle sizes.


\section*{Materials and Methods}

\subsection*{Materials}
Polyethylene glycol diacrylate (PEGDA 700, MW = 700 g/mol), lithium phenyl-2,4,6-trimethyl\-benzoylphosphinate (LAP), Span 80 and mineral oil were purchased from Sigma-Aldrich (St. Louis, MO).
Ultra-pure, deionized (DI) water (resistance = 18.2 M$\Omega$ cm) was used to disperse the polymers.
Square capillaries (0.10 mm $\times$ 1.0 mm, with a wall thickness of 0.09 mm) were purchased from Friedrich \& Dimmock Inc. (Millville, NJ). The ultraviolet (UV)-curable glue was purchased from Norland Optical Adhesive (Jamesburg, NJ).

Dulbecco's Modified Eagle Medium (DMEM, 1$\times$), fetal bovine serum (FBS), paraformaldehyde (PFA), phosphate-buffered saline (PBS), 0.25 wt\% Trypsin-EDTA (1$\times$), Pen Strep (100$\times$), and CellTracker (C34565) were purchased from Thermo Fisher Scientific (Waltham, MA). Phalloidin-iFluor488 (ab176753) was purchased from Abcam (Cambridge, United Kingdom). NIH3T3 cells were acquired from the Yale Cancer Center cell line repository as a frozen aliquot at an early passage (P$<$5). Passages P5 to 20 were used in this study. 

\subsection*{Sample preparation and video acquisition}

Yellow-green fluorescent polystyrene particles ($2a$ = 2 $\mu$m) were used as tracers. The tracers were gently mixed with the soft particles and introduced into a glass capillary square by capillary action. The soft particles were then allowed to settle to the bottom of the capillary. We tracked the motion of tracers near the bottom of the pseudo-2D system, which allowed us to determine the particle area fraction from wide-field images for each experiment. The square glass capillary was then sealed and affixed to a glass slide with optical glue to ensure stability and prevent unwanted evaporation and leakage during imaging.
The trajectory of the tracers was captured using an inverted fluorescent microscope (Zeiss Axio Observer 7, White Plains, NY) using the GFP channel (peak wavelength = 517 nm). The matrix particles are unstained and therefore do not show up in the fluorescence channel, but they can be captured using the bright field channel. Videos were recorded at a frame rate of 2 frames per second (lag time $\Delta t$ = 0.5 sec) for $T$ = 2000 frames.

\subsection*{Soft particle preparation}
Soft particles were prepared by photopolymerizing water droplets in a water-in-oil emulsion generated by vortex mixing. The process began with the preparation of two phases: an oil phase containing mineral oil and Span 80, a surfactant to stabilize the water-oil interface, and a water phase composed of deionized water, PEGDA 700, and LAP as a photoinitiator.
After preparing the solutions, the oil and water phases were combined in a glass vial at a volume ratio of 6:1 (oil phase : water phase) and vigorously mixed using a vortex mixer for 1 minute, producing a water-in-oil emulsion. The emulsion was then poured onto a petri dish and exposed to UV light (power = 6 W) for 1 minute to initiate photopolymerization, thereby curing the PEGDA 700 into hydrogel particles. The photo-crosslinked particles were collected in a conical tube, where they were washed by adding water and centrifuging at 2200 rpm for 15 minutes to separate the particles from the oil phase. Once the supernatant was decanted, the particles were resuspended in DI water. This washing procedure was repeated several times to ensure that we completely removed the residual oil and surfactant. 
Particle size was measured using Fiji (ImageJ) to obtain a mean radius of $\bar{R} = 4.35 \pm 1.31~\mu\mathrm{m}$.

\subsection*{Cell culturing and tracer introduction}

NIH3T3 cells were maintained in complete medium (89 vol\% DMEM, 10 vol\% FBS, and 1 vol\% Pen Strep) at 37~$^\circ$C and 5\% CO$_2$ and were subcultured every 2--3 days at 70--80\% confluency. Cells were lifted by adding Trypsin and seeded onto a 35 mm glass-bottom culture dish (MatTek, Ashland, MA) at a density of $\approx$100 mm$^{-2}$. Upon cell attachment, yellow-green fluorescent tracers ($2a$ = 100, 200 nm) were introduced at 0.02 vol\% into the complete media. The tracers ($2a$ = 100, 200 nm) were endocytosed by cells overnight. Cells were rinsed three times to remove free tracer particles prior to imaging. The imaging was performed while the dish was placed in an onstage incubator at 37 $^\circ$C perfused with 5\% CO$_2$. Videos were recorded at a frame rate of 33 frames per second (lag time $\Delta t$ = 30 ms) for $T$ = 2000 frames. To obtain the size distribution of the organelle structures, cells were fixed with 5\% PFA in PBS and stained with CellTracker and phalloidin according to the manufacturer's instructions.

\subsection*{Analysis procedures}

The captured videos of the tracers were analyzed using a multiple particle tracking (MPT) algorithm based on Ref. \cite{crocker1996methods}, developed in-house \cite{luo2022high,luo2025optimizing} and builds upon Refs. \cite{gao2009accurate,bayles2017probe}, to obtain the tracer coordinates $(x_{1,j}(t), x_{2,j}(t))$ at time point $t$ for the $j$-th tracer particle, where $i = 1,2$ denotes the two Cartesian directions. From each single-particle trajectory, we computed the time-averaged mean squared displacement (TAMSD) of the $j$-th particle as
\begin{equation}
    \Delta r_j^2(\Delta t) = \frac{1}{T - \Delta t} \sum_{t=0}^{T - \Delta t} \big[ (x_{1,j}(t + \Delta t) - x_{1,j}(t))^2 
   + (x_{2,j}(t + \Delta t) - x_{2,j}(t))^2 \big].
\end{equation}

The time- and ensemble-averaged 2D mean squared displacement (MSD) for each lag time $\Delta t$ was then calculated as
\begin{equation}
\langle \Delta r^2(\Delta t)\rangle = \frac{1}{N_t}\sum_{j=1}^{N_t}\Delta r_j^2(\Delta t),
\end{equation}
where $N_t$ is the number of tracer particles and $\langle \cdot \rangle$ denotes averaging over time and over the ensemble of particles.

To characterize not only the second moment of the motion but also the full displacement statistics, we additionally computed the self-part of the van Hove correlation function. For a given lag time $\Delta t$, the one-dimensional self-part along the $x$ direction is defined as
\begin{equation}
G_s(\Delta x,\Delta t)
=
\left\langle
\delta\!\left[
\Delta x - \bigl(x_{1,j}(t+\Delta t)-x_{1,j}(t)\bigr)
\right]
\right\rangle_{j,t},
\label{eq:van_hove}
\end{equation}
where $\delta(\cdot)$ is the Dirac delta function, and $\langle \cdot \rangle_{j,t}$ denotes averaging over all tracer particles and all time origins. Thus, $G_s(\Delta x,\Delta t)$ represents the probability distribution of single-particle displacements over the lag time $\Delta t$. In this work, we use $G_s(\Delta x,\Delta t)$ to compare the experimentally measured and simulated displacement distributions at fixed lag time, complementing the MSD analysis.

To verify the MSDs and extend the range of analyzed $\Delta t$, we also used  AIUQ. For AIUQ analysis, the time series was subsampled to include only every fourth frame ($\Delta t = 2$ sec, $T = 500$ frames). More details on tracking parameters for MPT can be found in Section 1 of the supplementary text.

\section*{Acknowledgements}
The authors thank Yuxin Luo for assistance with cell imaging.

\section*{Funding}
YL and JL are partially supported by the donors of the American Chemical Society Petroleum Research Fund under Doctoral New Investigator No. 67734-DNI10. MG acknowledges National Science Foundation Award OAC-2411043. YL acknowledges National Science Foundation Award OAC-2411044. 
JL is grateful for support from the Hyundai Motor Chung Mong-Koo Foundation and the Yale PEB Endowment.

\section*{Author contributions}
YL and JL designed the project. JL performed experiments and MC simulations. MG developed the PPGP predictive model for predicting the MSD over the parameter space, and the AIUQ approach for analyzing microscopy videos. JL and TL calibrated the PPGP model and performed the numerical analysis. All authors have contributed to the writing of the manuscript and approved the final version of the manuscript.

\section*{Conflicts of interest}
The authors declare no conflicts of interest. 

\section*{Data availability statement}
Codes generated as a part of this study have been made available through a GitHub repository: \url{https://github.com/jinseoklee97/tracer-diffusion-crowded-environments}


\newpage


\renewcommand{\thefigure}{S\arabic{figure}}
\renewcommand{\thetable}{S\arabic{table}}
\renewcommand{\theequation}{S\arabic{equation}}
\renewcommand{\thepage}{S\arabic{page}}
\setcounter{figure}{0}
\setcounter{table}{0}
\setcounter{equation}{0}
\setcounter{page}{1} 


\begin{center}
\section*{Supplementary Materials for ``Accessible pore geometry governs tracer diffusion in crowded environments''}

Jinseok Lee$^{1}$,
Tong Lin$^{2}$, Mengyang Gu$^{2}$,
Yimin Luo$^{1\ast}$\\ 
\small$^\ast$Corresponding author. Email: yimin.luo@yale.edu\\
\end{center}


\subsubsection*{This PDF file includes:}
Supplementary Text\\
Figures S1 to S10\\
Table S1\\
Captions for Movies S1 to S3\\

\newpage



\section*{Supplementary Text}

\section{Method Comparison}
\subsection{Multiple particle tracking (MPT)}

 MPT is widely regarded as a standard approach for analyzing tracer dynamics because it directly reconstructs individual particle trajectories. Important parameters include: 
\begin{itemize}
    \setlength{\itemsep}{0pt}
    \setlength{\parskip}{0pt}
\item \texttt{rad} - specifies the approximate radius of the object being tracked to allow filtering 
\item \texttt{memory} - specifies how many frames a feature is allowed to disappear from the tracking and then reemerge and still be considered part of the same trajectory 
\item \texttt{maxdisp} - specifies the maximum displacement a feature may make between successive frames. 
\item \texttt{Imin} - specifies the minimum brightness at the center of each feature. 
\end{itemize}

We find that the choice of these parameters, especially \texttt{maxdisp}, has a strong effect on the resulting MSD.  At small $\Delta t$'s, larger \texttt{maxdisp} is needed to capture the diffusive behavior. At large $\Delta t$'s, smaller \texttt{maxdisp} is needed to capture the more confined motion. In fact, we could not find a set of parameters that are consistent and applicable across all experimental $\Phi$ conditions. Furthermore, at longer lag times the number of available trajectory segments decreases substantially, which leads to increased statistical uncertainty in the MSD estimates. Consequently, it is common practice to consider only the first 10--20\% of the lag-time range as statistically reliable. 

\subsection{Ab initio uncertainty quantification (AIUQ) analysis}

In contrast, AIUQ applies the same analysis threshold across all datasets and does not require specifying any parameters. We first briefly summarize the mathematical framework of the estimation of MSD curves from AIUQ \cite{gu2024ab}, which assumes a parametric model of the intermediate scattering function. We extend the framework to obtain a model-free estimation \cite{lin2026model} of the MSD curves, used for analyzing experimental videos in this study. 


We first vectorize the intensities of a microscopy image with $N=N_1\times N_2$ pixels at each time frame $t$ and denote the vector by
\begin{equation}
    \mathbf y(t)=[y(\mathbf x_1,t),...,y(\mathbf x_{N},t)]^T,
\end{equation} where $\mathbf x_i=(x_{i,1},x_{i,2})^T$ is the 2D location of the pixel $i$ in the Cartesian space, for $i=1,...,N$. 

AIUQ imposes a probabilistic model of the image intensity in the original Cartesian space 
\begin{equation}
 \mathbf {y}(t)= \frac{1}{\sqrt{N} }\mathbf W^{*} \mathbf z(t) + \bm \epsilon(t), 
 \label{equ:SAM}
 \end{equation}
where  the  $N \times N$ matrix $\mathbf W^{*}$ is a 2D 
 complex conjugate of the Fourier basis, which relates the $N$ observations of an image at time $t$ from Cartesian space $\mathbf x=(x_1,x_2)^T$ to a vector of complex-valued random factor processes $\mathbf z(t)$ in the reciprocal space, where $\mathbf q=(q_1, q_2)^T$, $\bm \epsilon(t)\sim \mathcal{MN}(\mathbf 0, \frac{\bar B}{2} \mathbf I_N)$ is an N-dimensional Gaussian white noise vector with variance $\frac{\bar B}{2}$, and $\mathbf I_N$ is the identity matrix  of N dimensions. 
 
 The $N$-dimensional complex-valued latent factor $\mathbf z(t)$ is split into the real and imaginary parts: $\mathbf z(t)=\mathbf z_{re}(t)+i\mathbf z_{im}(t) $, where $\mathbf z_{re}(t)=(z_{1,re}(t),...,z_{N,re}(t))^T$ and $\mathbf z_{im}(t)=(z_{1,im}(t),...,z_{N,im}(t))^T$. 
We consider isotropic processes in this work, where the $j'$th row of the latent factor has the same intermediate scattering function (ISF) corresponding as the $j$th ring of the Fourier-transformed image intensity. For any $j'$, the random factor vectors over $n$ time points $\mathbf z_{\mathbf j',re}= (z_{\mathbf j',re}(t_1),...,z_{\mathbf j',re}(t_n))^T$ and $\mathbf z_{\mathbf j',im}= (z_{\mathbf j',im}(t_1),...,z_{\mathbf j',im}(t_n))^T$ are assumed to independently follow multivariate normal distributions 
 $\mathbf z_{\mathbf j',re}\sim \mathcal{MN}(\mathbf 0, \frac{A_j}{4}\mathbf R_j)$ and  $\mathbf z_{\mathbf j',im} \sim \mathcal{MN}(\mathbf 0, \frac{A_j}{4}\mathbf R_j)$, where $\frac{A_j}{4}\mathbf R_j$ is the covariance matrix for $j=1,...,J$ with $J$ being the total number of wave vector rings of the Fourier transformed image intensity. The $(k_1,k_2)$th entry of $\mathbf R_j$ is characterized by ISF: $R_j(k_1,k_2)= f_{\bm \theta}(q_j, \Delta t_k)$ with $\Delta t_k=|k_2-k_1| \Delta t_{min}$ with $\Delta t_{min}$ being the interval between two consecutive time  frames and $f_{\bm \theta}(q_j, \Delta t_k)$ is the ISF with a vector parameters $\bm \theta$ at a scalar Fourier magnitude $q_j$  and lag time point $\Delta t_k$, for $j=1,...,J$ and   $\Delta t_k$. 

Instead of assuming a parametric model of the ISF, we approximate the ISF from the MSD by the cumulant theorem \cite{koppel1972analysis,lin2026model}:
\begin{equation}
f_{\boldsymbol{\theta}}(\mathbf{q}, \Delta t)
\approx
\exp\left(-\frac{q^2 \theta(\Delta t)}{4}\right),
\end{equation}
where \(f_{\boldsymbol{\theta}}(\mathbf{q}, \Delta t)\) denotes the ISF, \(\theta(\Delta t)\) denotes the MSD at lag time \(\Delta t\), and \(q = \|\mathbf{q}\|\) is the magnitude of the wave vector \(\mathbf{q}\). This approximation removes the need for a closed-form ISF model, thereby enabling model-free estimation of the MSD. Instead of using direct inversion approaches \cite{bayles2017probe,gu2021uncertainty}, we estimate the MSD curves by the maximum marginal likelihood estimator (MMLE)  \cite{gu2024ab}, which naturally weighs information across different frequencies. As the MMLE borrow informations across the time domain in estimating MSDs, the estimated MSDs are available throughout the full temporal range, while previous model-free approaches can only produce model-free estimation at several lag times in some scenarios  \cite{bayles2017probe,gu2021uncertainty}. 

\begin{figure}[htb!]
\centering
\includegraphics[width=0.90\textwidth]{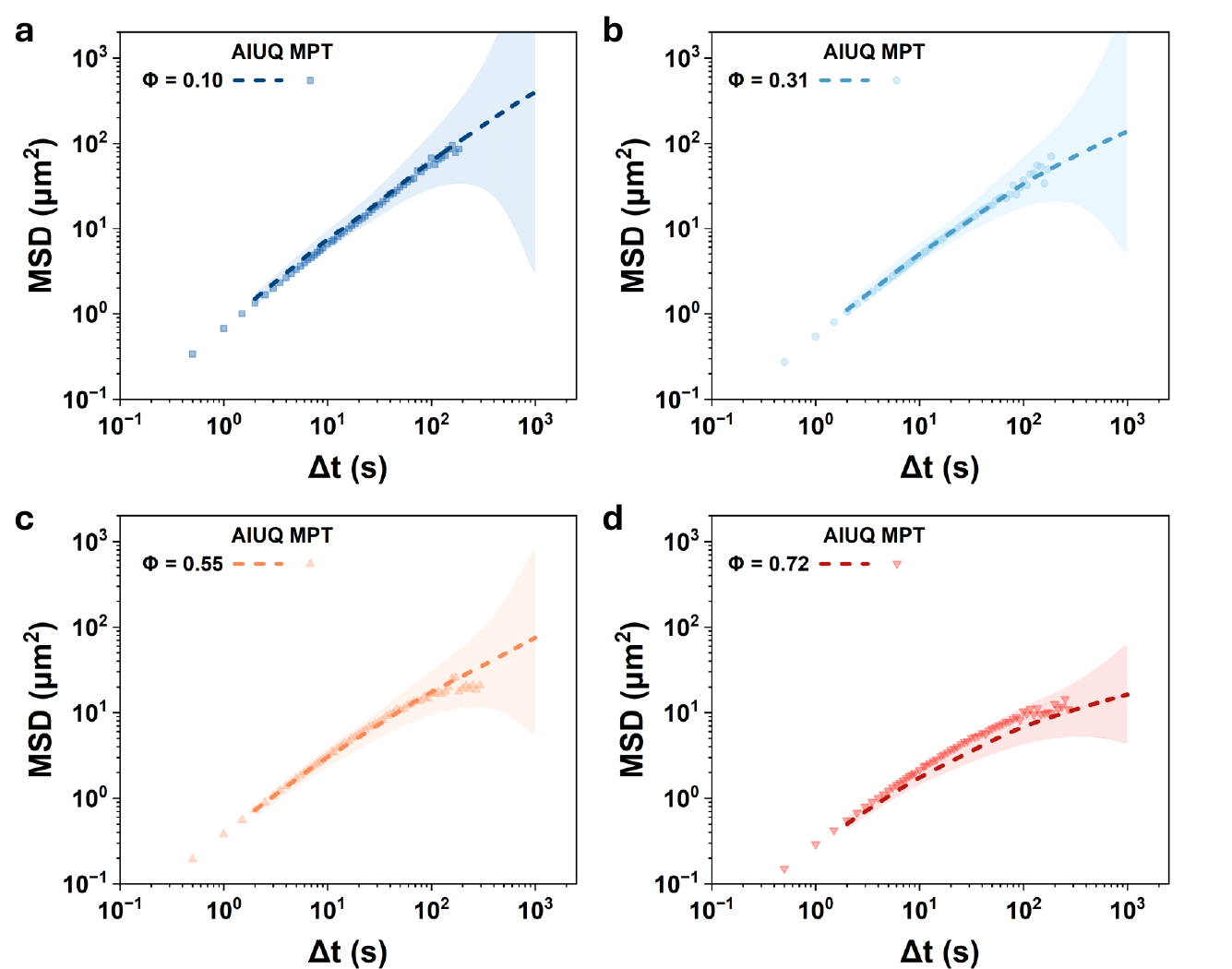}
   \caption{\textbf{Agreement between AIUQ and MPT.}
   Experimental MSDs analyzed using both MPT 
   (symbols) and AIUQ (dashed lines). The shaded regions represent the 95\% predictive interval from AIUQ.
   (a) $\Phi=0.10$, $\bar R=3.30$, $p=0.398$.
   (b) $\Phi=0.31$, $\bar R=3.86$, $p=0.373$.
   (c) $\Phi=0.55$, $\bar R=3.73$, $p=0.323$.
   (d) $\Phi=0.72$, $\bar R=4.56$, $p=0.286$.
   }
\label{fig:exp_AIUQ_pred}
\end{figure}

 We find that the resulting MSDs agree well with the optimized MPT parameter results while remaining consistent across different experimental conditions (Fig. \ref{fig:exp_AIUQ_pred}).

\section{Simulation procedures}

We develop a minimal model for tracer transport in crowded environments. 
It serves three purposes. 
First, it reproduces the tracer dynamics observed in the soft particle experiments using 
experimentally measured geometric parameters, including the matrix area fraction $\Phi$, 
mean particle radius $\bar{R}$, and polydispersity $p$. Second, it enables systematic exploration of the parameter space that governs tracer transport, 
which would be impractical to probe experimentally due to the substantial time required for 
sample preparation, imaging, and analysis. 
Third, the simulation results enable the inference of structural properties of intracellular 
environments, such as the effective area fraction and the accessible pore size.

To this end, we perform time-series simulations of the motion of $N_t$ tracer particles and 
$N_m$ matrix particles, from $t=\Delta t$ to $t=\Delta t \times T$ with a time step $\Delta t$. 
The position of the $k$-th matrix particle is denoted by 
$\mathbf{X}_k(t) = (X_{1,k}(t), X_{2,k}(t))$, and that of the $j$-th tracer particle by 
$\mathbf{x}_j(t) = (x_{1,j}(t), x_{2,j}(t))$.

All simulation parameters used in this study are summarized in Table~\ref{tab:sim_params}. Two parameter sets are used: a generic parameter set used to reproduce the soft particle experiments and explore the parameter space, and an intracellular parameter set used to model tracer transport within living cells.

\subsection{Initialization: Filling the space with matrix particles}

We initialize the simulation by placing the matrix particles in a 2D, $L \times L$ simulation box. The number of matrix particles $N_m$ is constrained by geometric parameters. To construct a simulation box containing matrix particles with a target area fraction $\Phi = \tfrac{1}{L^2}\sum_{k=1}^{N_m}\pi R_k^2$, we first generat an array of polydisperse particles whose radii are sampled from a normal distribution with mean radius $\bar{R} = \tfrac{1}{N_m}\sum_{k=1}^{N_m} R_k$ and polydispersity $p=\sqrt{\tfrac{1}{N_m}\sum_{k=1}^{N_m}(R_k-\bar R)^2}\bigr/(\bar R)$. Particles are then sequentially placed in the box in descending order of size. The placement algorithm consists of two main steps:

\subsubsection{Random insertion} 
For the $k$-th matrix particle of radius $R_k$, a trial position $\mathbf X_k=(X_{1,k}, X_{2,k})$ is randomly selected from the domain $[R_k, L - R_k]^2$, ensuring the particle lies fully within the square box of side length $L$. The position is accepted only if it does not overlap with any previously placed particle $j < k$, i.e.,
\begin{equation}
(X_{1,k} - X_{1,j})^2 + (X_{2,k} - X_{2,j})^2 > (R_k + R_j)^2 \quad \text{for all } j < k.
\end{equation}
If no valid position is found after 10,000 attempts, the algorithm transitions to a local relaxation scheme.

\subsubsection{Local relaxation}
When random insertion fails, a local relaxation procedure is performed:

\paragraph{Brownian random walk.}
Each matrix particle $k$ is perturbed by a stochastic displacement:
\begin{equation}
\mathbf{X}_k \leftarrow \mathbf{X}_k + \sigma\, \boldsymbol{\xi}_k, \quad \boldsymbol{\xi}_k \sim \mathcal{MN}(\mathbf 0, \mathbf{I}_2),  
\end{equation}
where $\sigma = \sqrt{2 D_{\text{m}} \Delta t}$ is the thermal fluctuation scale, $D_{\text{m}}$ is the Brownian diffusivity of the matrix particles and $\mathcal{MN}$ denotes a multivariate normal distribution.

\paragraph{Pairwise repulsion.}
After the random walk step, overlapping particle pairs are identified and a pairwise repulsive force is applied:
\begin{equation}
\mathbf{F}_{kj} = k_{\text{rep}} \delta_{kj} \hat{\mathbf{r}}_{kj},   
\end{equation}
\begin{equation}
\delta_{kj} = R_k + R_j - r_{kj}, \quad \text{if } \delta_{kj} > 0,  
\end{equation}
where $r_{kj} = \|\mathbf{X}_k - \mathbf{X}_j\|$ is the inter-particle distance and $\hat{\mathbf{r}}_{kj}$ is the unit vector from particle $j$ to particle $k$. The net force $\mathbf{F}_k$ acting on each particle is then used to update its position according to the discretized overdamped Langevin equation (with $\gamma = 1$):
\begin{equation}
\mathbf{X}_k \leftarrow \mathbf{X}_k + \mathbf{F}_k \Delta t.    
\end{equation}

\begin{figure}
\centering
\includegraphics[width=\textwidth]{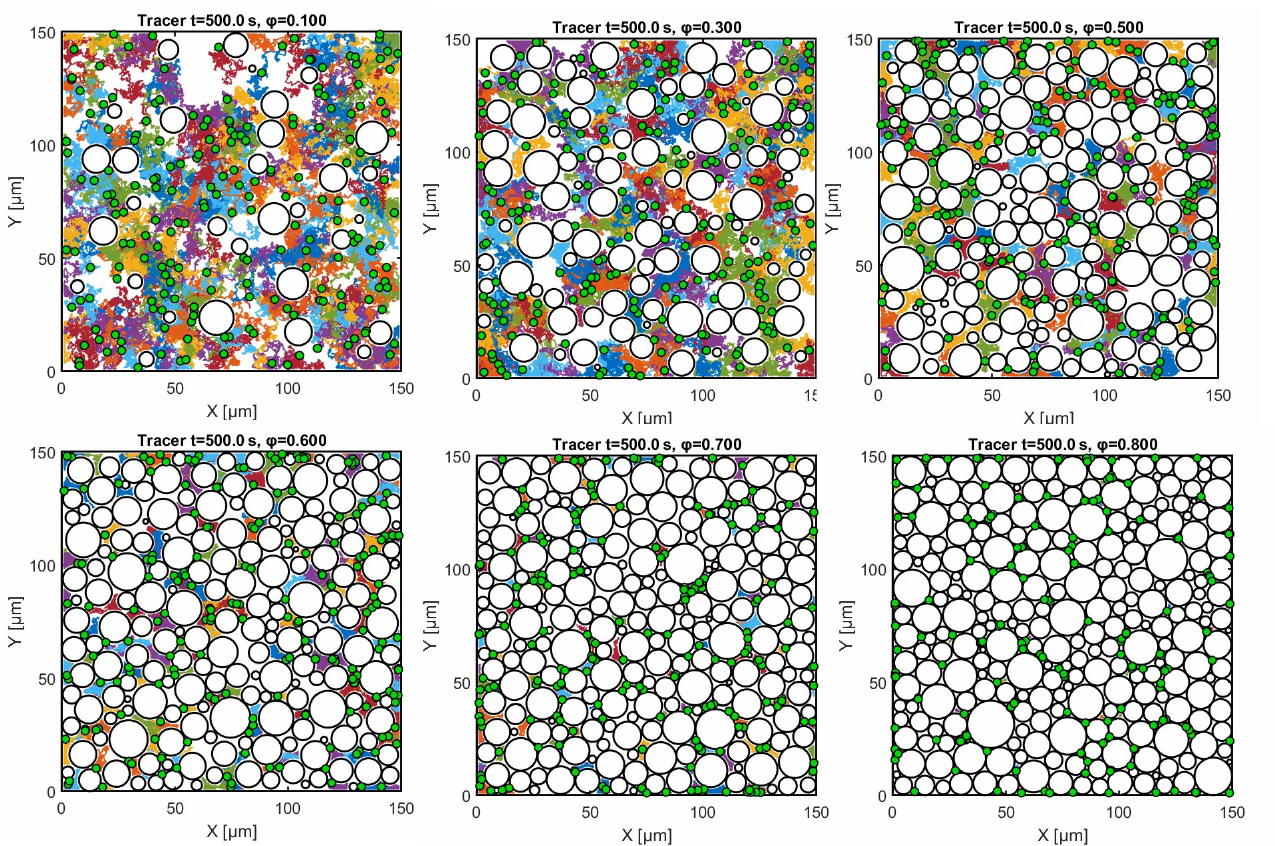}
   \caption{\textbf{Initial configurations of matrix particles at different area fractions.}
   Matrix particles are shown as white circles, and tracer trajectories are overlaid as colored lines.}
\label{fig:simulation_space_filling}
\end{figure}

\begin{figure}
\centering
\includegraphics[width=0.6\textwidth]{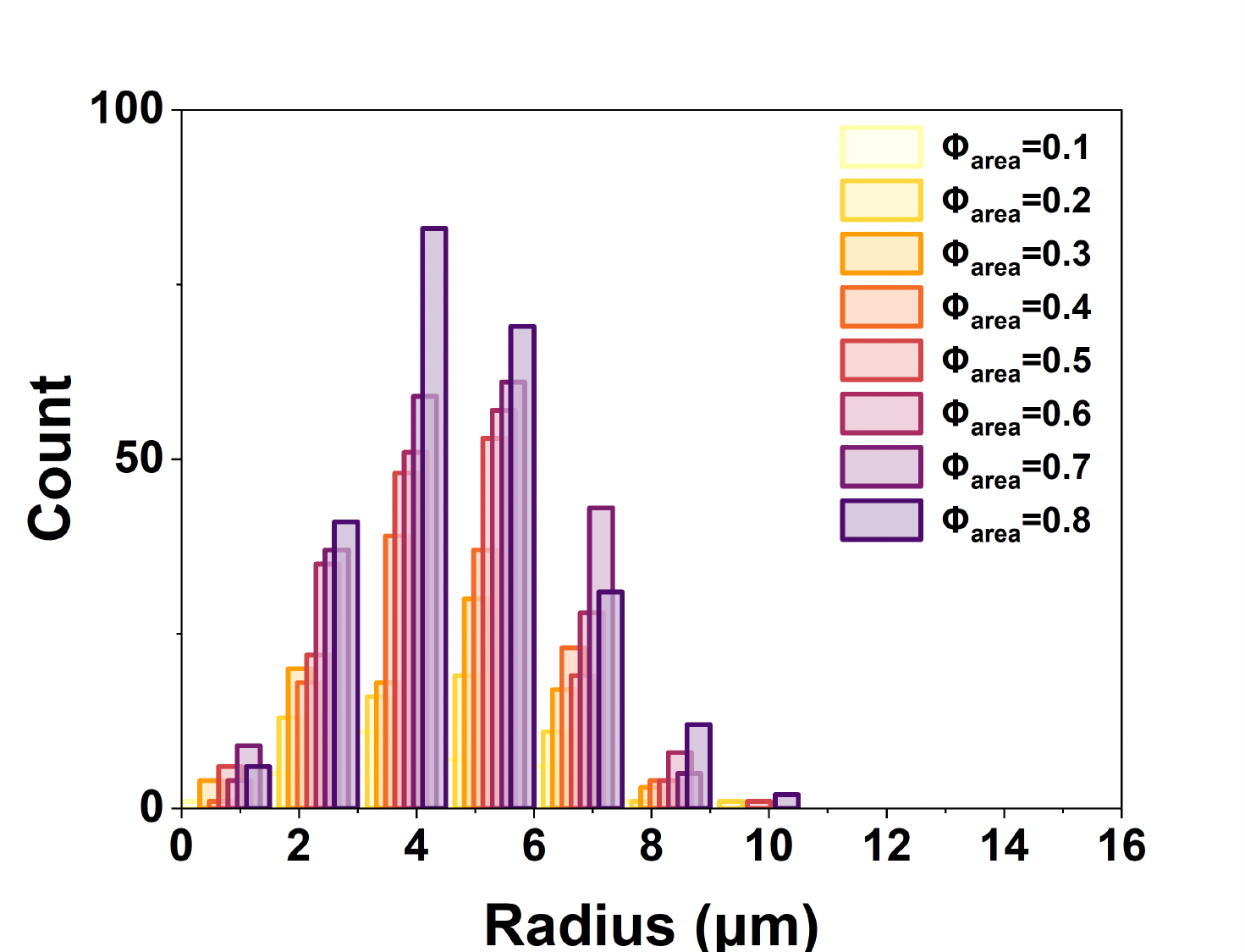}
   \caption{\textbf{Matrix particle size distribution in simulation.}
   The distribution is identical across simulations at different area fractions.}
\label{fig:matrix_size_distr_diff_vol_fraction}
\end{figure}

\paragraph{Geometric correction.}
If residual overlaps remain even after repulsion, they are corrected by displacing each overlapping pair along the line of centers by $\pm \delta_{kj}/2$, ensuring exact contact:
\begin{equation}
\mathbf{X}_k \leftarrow \mathbf{X}_k + \frac{\delta_{kj}}{2} \hat{\mathbf{r}}_{kj}, \quad
\mathbf{X}_j \leftarrow \mathbf{X}_j - \frac{\delta_{kj}}{2} \hat{\mathbf{r}}_{kj}. 
\end{equation}
The initial state of the matrix particles in different area fractions after filling the space is shown in Fig.~\ref{fig:simulation_space_filling}, with the matrix particle size distribution shown in Fig.~\ref{fig:matrix_size_distr_diff_vol_fraction}, where the matrix particle size distribution remains consistent across all area fractions.
A representative matrix particle packing process is shown in Supplementary Movie S2.

\subsection{Running the simulation: Position updates}

At each simulation step, the positions of tracer and matrix particles evolve according to overdamped Brownian dynamics while enforcing hard-sphere exclusion and confinement within the square domain. Tracer trajectories represent stochastic diffusion constrained by steric interactions with the surrounding matrix and by hydrodynamic hindrance arising from nearby boundaries and obstacles. Matrix particles undergo Brownian motion combined with relaxation steps that eliminate overlaps and preserve the target packing structure. We distinguish the update rules for the tracers and matrices and the correction applied when the motion of the matrix induces residual overlaps.

\subsubsection{Tracer particles}  
In a pure solvent, the Brownian motion of a tracer particle is governed by the bulk viscosity of the medium and the particle’s own size, resulting in a global diffusion coefficient \(D_{\mathrm{0}}\). In our experimental system, however, tracers move in close proximity to other objects, such as matrix particles and the bottom wall. As a tracer approaches an interface or another particle, its diffusivity is reduced due to hydrodynamic hindrance arising from confinement and near-field interactions \cite{Bevan2000, brenner1961slow}. To account for this effect, the local diffusivity in the simulation is adjusted at each time step by a multiplicative hydrodynamic hindrance factor. The instantaneous in-plane diffusivity is defined as  
\begin{equation}
    D_{\mathrm{loc}} = f_{\mathrm{loc}}\, D_{\mathrm{0}},
    \qquad
    f_{\mathrm{loc}} = f_{\mathrm{wall}}\, f_{\mathrm{mat}} \in (0,1],
\end{equation}
where \(D_{\mathrm{loc}}\) is the local diffusion coefficient, \(f_{\mathrm{loc}}\) is the combined hydrodynamic hindrance factor, and \(f_{\mathrm{wall}}\) and \(f_{\mathrm{mat}}\) represent the contributions from the bottom wall and neighboring matrix particles, respectively.

\paragraph{Wall hindrance.}

\begin{figure*}
\centering
\includegraphics[width=0.75\textwidth]{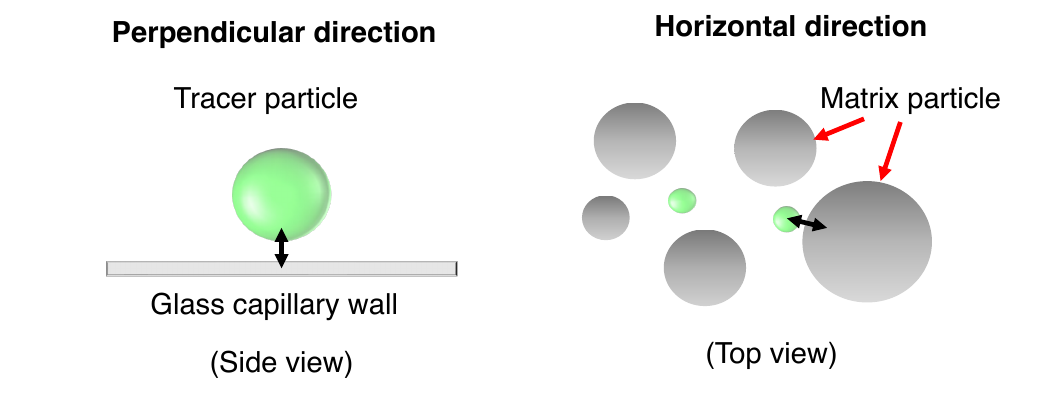}
   \caption{
   Schematic illustration of hydrodynamic drag corrections applied when tracers move above the glass capillary wall (left) and near matrix particles (right). 
}
\label{fig:wall_hindrance}
\end{figure*}

The reduction in tracer mobility for motion parallel to a planar wall (Fig. \ref{fig:wall_hindrance}, left) is approximated by the Faxén series~\cite{goldman1967slow}:
\begin{equation}
    f_{\mathrm{wall}}(h)
    = 1
    - \frac{9}{16}\lambda
    + \frac{1}{8}\lambda^{3}
    - \frac{45}{256}\lambda^{4}
    - \frac{1}{16}\lambda^{5},
    \qquad
    \lambda \equiv \frac{r_s}{h},
\end{equation}
where \(r_s\) is the tracer radius and \(h(t)\) is the instantaneous center height of the tracer above the wall. 
Although the simulation is performed in two dimensions, a notional height \(h(t)\) is introduced to capture the hydrodynamic coupling with the substrate. 
In experiments, tracer particles do not remain in direct contact with the bottom glass surface but fluctuate vertically due to a balance among gravitational, electrostatic, and thermal energies. 
To represent this weak confinement, the tracer height \(h(t)\) is modeled as an Ornstein–Uhlenbeck (OU) process with a reflecting lower boundary at \(h=r_s\):
\begin{equation}
    h(t+\Delta t)
    = h_0 + \bigl(h(t) - h_0\bigr)e^{-\Delta t/\tau_h}
      + \sigma_h \sqrt{1 - e^{-2\Delta t/\tau_h}}\,\xi_t,
    \qquad \xi_t \sim \mathcal{N}(0,1),
\end{equation}
where \(h_0 = 1.25\,r_s\) is the mean center height of the tracer, \(\tau_h\) is the OU relaxation time,  \(\sigma_h\) controls the amplitude of height fluctuations and $\mathcal N$ denotes a normal distribution. 

The Faxén expression assumes a no-slip boundary condition at the wall; however, in practice, the hindrance can be partially mitigated by factors such as electrostatic repulsion or surface mobility. 
To account for this, the wall factor is optionally blended toward unity by a coupling coefficient \(0 \le \beta \le 1\):
\begin{equation}
    f_{\mathrm{wall}} \leftarrow 1 - \beta\bigl(1 - f_{\mathrm{wall}}\bigr),
\end{equation}
where \(\beta=1\) corresponds to a fully no-slip wall and \(\beta=0\) corresponds to the absence of wall hindrance. 
The value of \(\beta\) is determined by comparing the simulated mean-squared displacement of tracers in the absence of matrix particles with the two-dimensional Einstein–Smoluchowski relation,
\begin{equation}
    \langle r^2(\Delta t) \rangle = 4 D \Delta t.
\end{equation}
\paragraph{Matrix particle hindrance.}
The hydrodynamic hindrance exerted by larger particles on tracers (Fig. \ref{fig:wall_hindrance}, right) is
known to increase sharply as the inter-particle gap decreases, since
the lubrication resistance diverges as the gap thickness approaches
zero, leading to a strong reduction in particle mobility near contact
\cite{KimKarrila1991}. Accordingly, the lubrication-type hindrance factor was modeled as a smooth, bounded function of the nearest surface-to-surface gap.
\begin{equation}
    f_{\mathrm{mat}}(g^\ast) \;=\; \left(\frac{g^\ast}{g^\ast + g_0}\right)^{\alpha_{\mathrm{hyd}}}, 
    \qquad 0<\alpha_{\mathrm{hyd}}\leq 1,
\end{equation}
where \(g^\ast\) is the surface-to-surface gap between the tracer and nearest matrix particle, \(g_0\) is a characteristic length scale that sets the onset of hydrodynamic hindrance, and \(\alpha_{\mathrm{hyd}}\) is an empirical exponent controlling the sharpness of the near-contact slowdown.

\paragraph{Step update.}
At each time step, the tracer displacement is drawn from a Gaussian distribution representing overdamped Brownian motion with the local diffusivity:
\begin{equation}
    \Delta \mathbf{x}_j(t) \sim \mathcal{MN}\!\bigl(\mathbf{0},\, 2 D_{\mathrm{loc}} \Delta t\, \mathbf{I}_2\bigr),
\end{equation}
where \(\mathbf{I}\) is the identity tensor. 
Reflective boundary conditions confine the motion within the square domain. 
If a proposed displacement would result in overlap with any matrix particle,
\begin{equation}
    \bigl\|\mathbf{x}_j(t) + \Delta \mathbf{x}_j(t) - \mathbf{X}_k(t)\bigr\| < r_s + R_k,
\end{equation}
the step is rejected and resampled until a non-overlapping configuration is obtained. 
The accepted update is then applied as
\begin{equation}
    \mathbf{x}_j(t+\Delta t) = \mathbf{x}_j(t) + \Delta \mathbf{x}_j(t).
\end{equation}

\subsubsection{Matrix particles}
Matrix particles are modeled as Brownian disks with reduced mobility relative to the tracers. 
In the soft particle experiments, the matrix particles are effectively immobilized, whereas 
in intracellular environments the matrix components may undergo slow rearrangements. 
To account for these differences, the matrix displacement is defined as

\begin{equation}
\Delta \mathbf{X}_k(t) \sim \mathcal{MN}\bigl(\mathbf{0},\,2D_{\text{m}}\,m\,\Delta t\,\mathbf{I}_2\bigr),
\end{equation}
where $m$ denotes the matrix mobility factor controlling 
the reduced Brownian motion of the matrix particles.

Reflective-wall conditions prevent escape across the boundaries. After each stochastic update, 
overlaps between matrix particles are resolved through a relaxation scheme: (i) overlapping pairs 
experience a linear repulsive force proportional to their overlap distance, (ii) positions are 
updated according to the overdamped Langevin equation, and (iii) any remaining overlaps are 
corrected geometrically by shifting each pair along their line of centers until they are in exact 
contact. This procedure ensures that the matrix configuration always satisfies the hard-sphere 
exclusion.

\subsubsection{Residual overlap removal}  
Although tracer steps are resampled until they do not overlap with the instantaneous matrix configuration, overlaps may still arise when matrix particles subsequently move. In such cases, any tracer that is found inside a matrix particle is displaced radially outward along the line of centers until it lies exactly at contact,
\begin{equation}
\mathbf{x}_j(t+\Delta t) \;=\; \mathbf{X}_k(t+\Delta t) + \frac{\mathbf{x}_j(t+\Delta t) - \mathbf{X}_k(t+\Delta t)}{\|\mathbf{x}_j(t+\Delta t) - \mathbf{X}_k(t+\Delta t)\|}\,(R_k + r_s + \epsilon),
\end{equation}
where $R_k$ and $r_s$ are the radii of the matrix and tracer particles, respectively, and $\epsilon$ is a small buffer to avoid numerical re-overlap. This post-update correction guarantees that hard-sphere exclusion is maintained even when matrix motion generates new contacts with tracers.

\begin{table}
\centering
\caption{\textbf{Simulation inputs and fixed settings.}
Two simulation classes are considered: a generic simulation class used to reproduce the soft-particle experiments and explore the geometric input space, and an intracellular simulation class used to model tracer transport in living cells. The swept input variables define the sampled geometric parameter space, whereas the remaining quantities are fixed within each simulation.}
\label{tab:sim_params}

\begin{tabular}{lcccc}
\hline
Quantity & Symbol & Generic simulations & Intracellular simulations & Unit \\
\hline

\multicolumn{5}{c}{\textbf{Input variables}} \\
\hline

Matrix area fraction & $\Phi$ & 0--0.75 & 0--0.80 & -- \\

Mean matrix radius & $\bar{R}$ & 2--6 & 0.4--1.2 & $\mu$m \\

Matrix polydispersity & $p$ & 0--1 & 0--1 & -- \\

\hline
\multicolumn{5}{c}{\textbf{Fixed simulation parameters}} \\
\hline

Matrix mobility factor & $m$ & 0 & 1 & -- \\

Tracer radius & $R_{\mathrm{tr}}$ & 1.0 & 0.05, 0.10 & $\mu$m \\

Simulation box size & $L$ & 150 & 20 & $\mu$m \\

Solvent viscosity & $\eta$ & $10^{-3}$ & $10^{-1}$ & $\mathrm{Pa\cdot s}$ \\

Tracer time step & $\Delta t$ & 0.5 & 0.015 & s \\

Number of time steps & $N_{\mathrm{steps}}$ & 2000 & 4000 & -- \\

Number of tracers & $N_{\mathrm{tr}}$ & 200 & 100 & -- \\

Mean tracer height & $h_0$ & $1.25R_{\mathrm{tr}}$ & $1.25R_{\mathrm{tr}}$ & $\mu$m \\

Height relaxation time & $\tau_h$ & 2.0 & 2.0 & s \\

Height fluctuation amplitude & $\sigma_h$ & $0.04R_{\mathrm{tr}}$ & $0.04R_{\mathrm{tr}}$ & $\mu$m \\

Wall-hindrance blending factor & $\beta$ & 0.4 & 0.4 & -- \\

Gap-response exponent & $\alpha_{\mathrm{hyd}}$ & 0.6 & 0.6 & -- \\

Gap length scale & $g_0$ & $0.30R_{\mathrm{tr}}$ & $0.30R_{\mathrm{tr}}$ & $\mu$m \\

Repulsive spring constant & $k_{\mathrm{repulse}}$ & 500 & 500 & -- \\

\hline
\end{tabular}
\end{table}

\section{Calculation of the non-Gaussian parameter ($\alpha_2$) and velocity autocorrelation function (VACF)}

\begin{figure}
\centering
\includegraphics[width=0.9\textwidth]{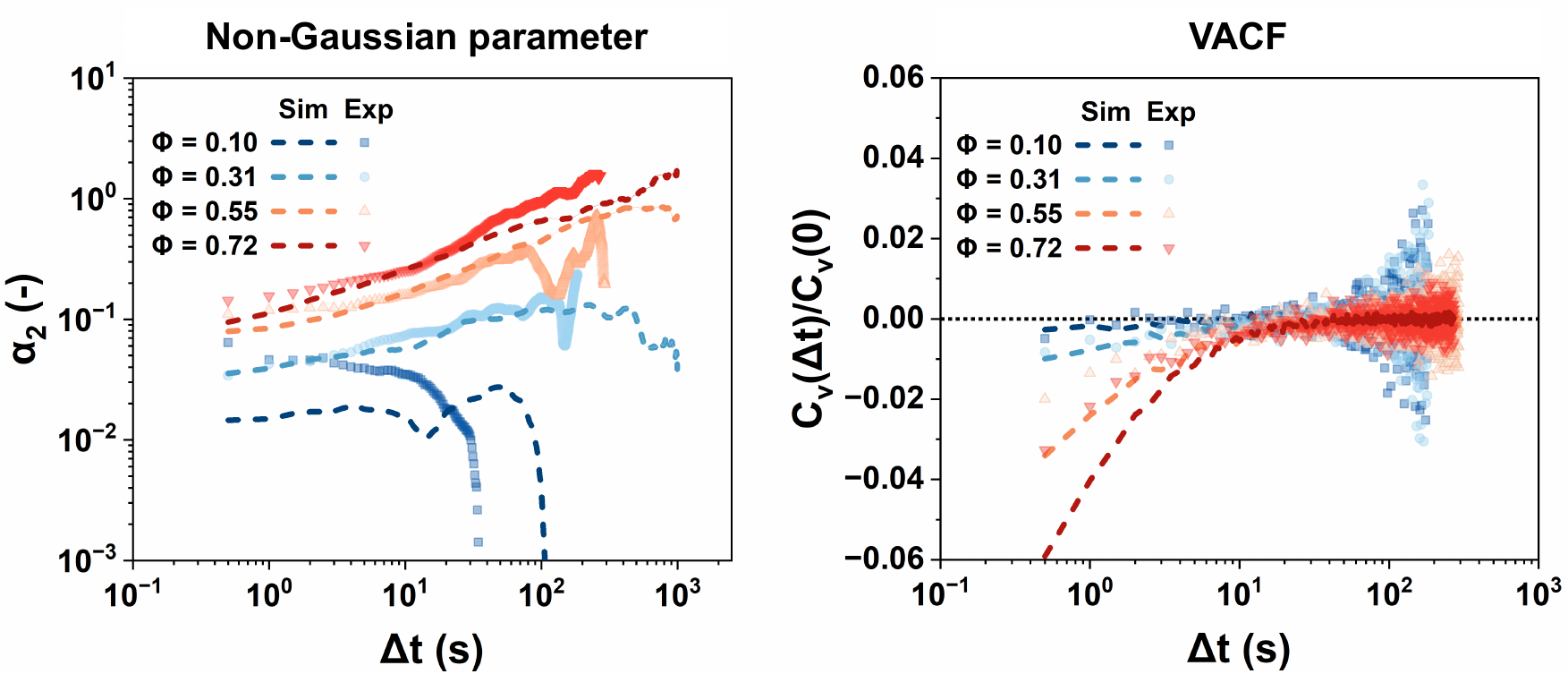}
   \caption{Comparison of the non-Gaussian parameter and velocity autocorrelation function between experimental and simulation.}
\label{fig:non_Gaussian_param_VACF}
\end{figure}

The non-Gaussian parameter was calculated from the displacement distribution at each lag time $\Delta t$ (Fig. \ref{fig:non_Gaussian_param_VACF}a). For each tracer and starting time, the displacement was defined as $\Delta \mathbf{r}(t,\Delta t)=\mathbf{r}(t+\Delta t)-\mathbf{r}(t)$. The radial non-Gaussian parameter was then calculated as
\begin{equation}
\alpha_2(\Delta t)=\frac{\langle \Delta r^4\rangle}{2\langle \Delta r^2\rangle^2}-1,
\end{equation}

where the brackets denote an average over all tracers and all valid starting times.

The velocity autocorrelation function (VACF, Fig. \ref{fig:non_Gaussian_param_VACF}b) was calculated from the frame-to-frame tracer velocity, $\mathbf{v}(t)=[\mathbf{r}(t+\delta t)-\mathbf{r}(t)]/\delta t$, where $\delta t$ is the imaging frame interval. For each lag time, the VACF was obtained as
\begin{equation}
C_{v}(\Delta t)=\langle \mathbf{v}(t)\cdot\mathbf{v}(t+\Delta t)\rangle.
\end{equation}

The reported VACF was normalized by its zero-lag value, $C_{v}(\Delta t)/C_{v}(0)$.

\section{Validation metrics for the PPGP predictive model}

To evaluate the predictive accuracy of the PPGP  model for the MSD scaling exponent $\alpha$, we use the root-mean-square error (RMSE), mean absolute error (MAE), and coefficient of determination ($R^2$). RMSE measures the typical magnitude of the prediction error while giving greater weight to larger deviations. MAE measures the average absolute difference between predicted and simulated values and is less sensitive to outliers than RMSE. The coefficient of determination $R^2$ quantifies how well the predicted values explain the variance in the simulated data.

\begin{equation}
\mathrm{RMSE} =
\sqrt{\frac{1}{N}\sum_{i=1}^{N}(\alpha_i-\hat{\alpha}_i)^2}
\end{equation}

\begin{equation}
\mathrm{MAE} =
\frac{1}{N}\sum_{i=1}^{N}|\alpha_i-\hat{\alpha}_i|
\end{equation}

\begin{equation}
R^2 =
1-\frac{\sum_{i=1}^{N}(\alpha_i-\hat{\alpha}_i)^2}
{\sum_{i=1}^{N}(\alpha_i-\bar{\alpha})^2}
\end{equation}

Here, $\alpha_i$ denotes the value obtained directly from simulation for the $i$-th test sample, $\hat{\alpha}_i$ is the corresponding prediction by the PPGP model, $\bar{\alpha}$ is the mean of the simulated $\alpha$ values over the test set, and $N=100$ is the number of held-out test samples used in the metric calculation.

\section{Procedures for obtaining percolation transition $\Phi_c$}

\begin{figure}[htb!]
\centering
\includegraphics[width=\textwidth]{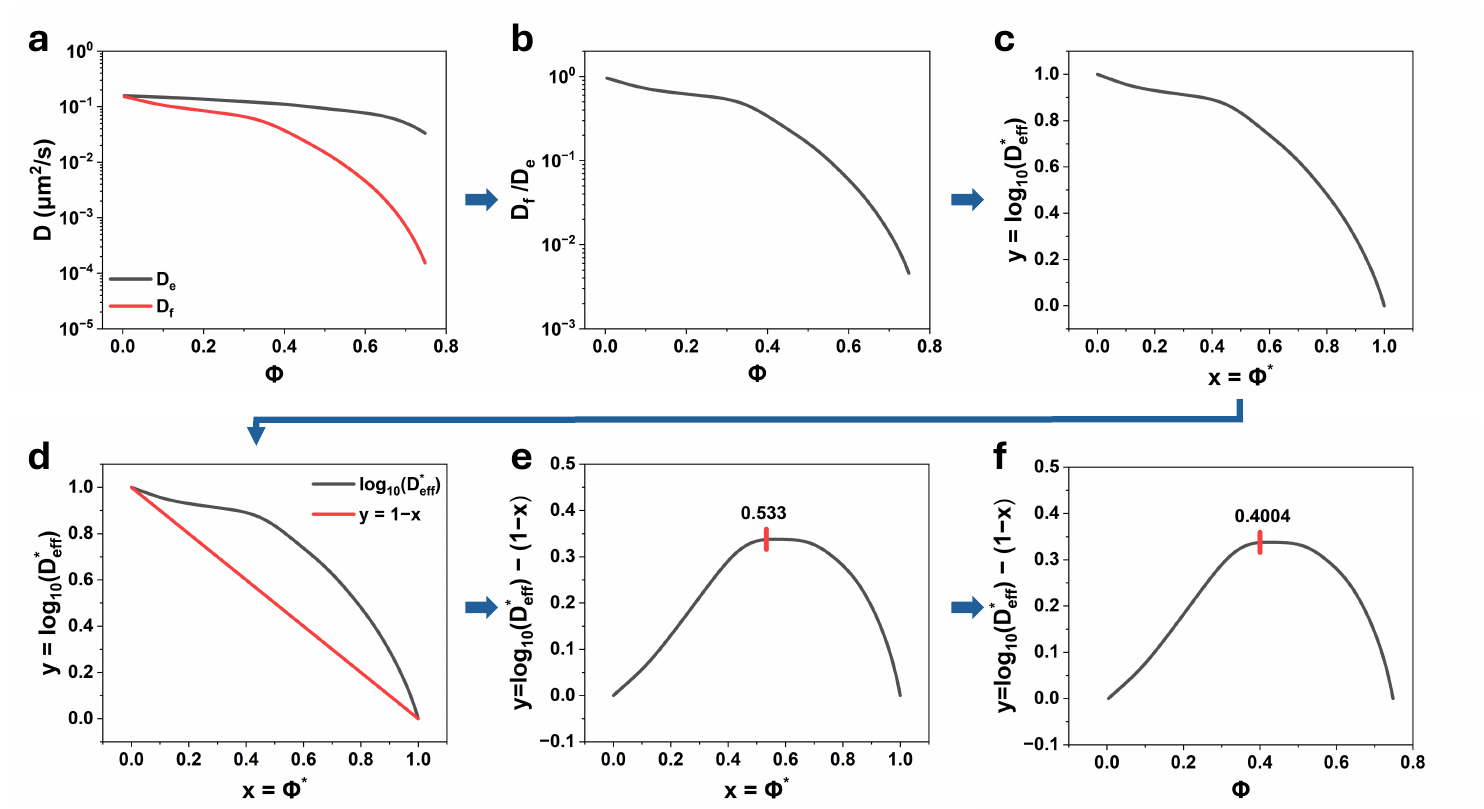}
   \caption{\textbf{Knee-point identification procedure.} (a-c) Transformation applied to highlight the knee point, indicating a ``phase change'' in the data.  
   (d-e) The knee is identified as the point of maximum deviation from the red reference line $f(x) = (1 - x)$.  
   (f) The data is transformed back to the original variable $\Phi$. }
\label{fig:finding_knee}
\end{figure}

In this work, we define the percolation threshold $\Phi_c$ to be the knee, a point where the curve visibly bends, of the quantity $D_\mathrm{eff}(\Phi)=\frac{D_\mathrm{f}(\Phi)}{D_\mathrm{e}(\Phi)}$. Here, ${D_\mathrm{e}(\Phi)}$ denotes the initial diffusion coefficient, obtained from the slope of the MSD versus $\Delta t$ curve over the range $\Delta t = 1$ to $10$. In contrast, ${D_f(\Phi)}$ represents the final diffusion coefficient, calculated by averaging the MSD versus $\Delta t$ data over the interval $\Delta t = 500$ to $1000$ (Fig.~\ref{fig:finding_knee}a). This procedure is illustrated in finding $\Phi_c$ for composition $\bar{R}$ = 3.76 and $p$ = 0.500. To eliminate the effect of initial confinement on the diffusion coefficient, we take the ratio of ${D_\mathrm{f}}$ to ${D_\mathrm{e}}$ (Fig.~\ref{fig:finding_knee}c) to find ${D_\mathrm{eff}}$ (Fig.~\ref{fig:finding_knee}b). To quantitatively determine the knee location, we employ the \textit{kneedle} algorithm. First, both the area fraction $\Phi$ and $\log_{10} D_{\mathrm{eff}}$ are normalized to the interval $[0,1]$:
\begin{align}
x&=\Phi^{*}
=
\frac{\Phi - \Phi_{\min}}{\Phi_{\max} - \Phi_{\min}}
\\[6pt]
f(x)& = \log_{10} D_{\mathrm{eff}}^{*}
=
\frac{
\log_{10} D_{\mathrm{eff}} - \log_{10} D_{\mathrm{eff}}^{\min}
}{
\log_{10} D_{\mathrm{eff}}^{\max} - \log_{10} D_{\mathrm{eff}}^{\min}
}
\end{align}

We then construct the reference line $
f(x)= 1-x$ (Fig.~\ref{fig:finding_knee}d), and compute the difference between this line and the normalized curve. The knee point $\Phi_c^{*}$ is identified as the value of $x$ at which this difference attains its maximum (Fig.~\ref{fig:finding_knee}e). For the representative case shown here, we obtain $\Phi_c^{*}=0.533$. Finally, the percolation threshold $\Phi_c$ in the original scale is recovered by inverse normalization: $\Phi_{\mathrm{c}}
= \Phi_{\mathrm{c}}^{*} \times (\Phi_{\max} - \Phi_{\min}) + \Phi_{\min} = 0.4004$.

\section{Pore-size characterization}
\subsection{Experimental measurement of the pore size distribution}
\begin{figure}
\centering
\includegraphics[width=\textwidth]{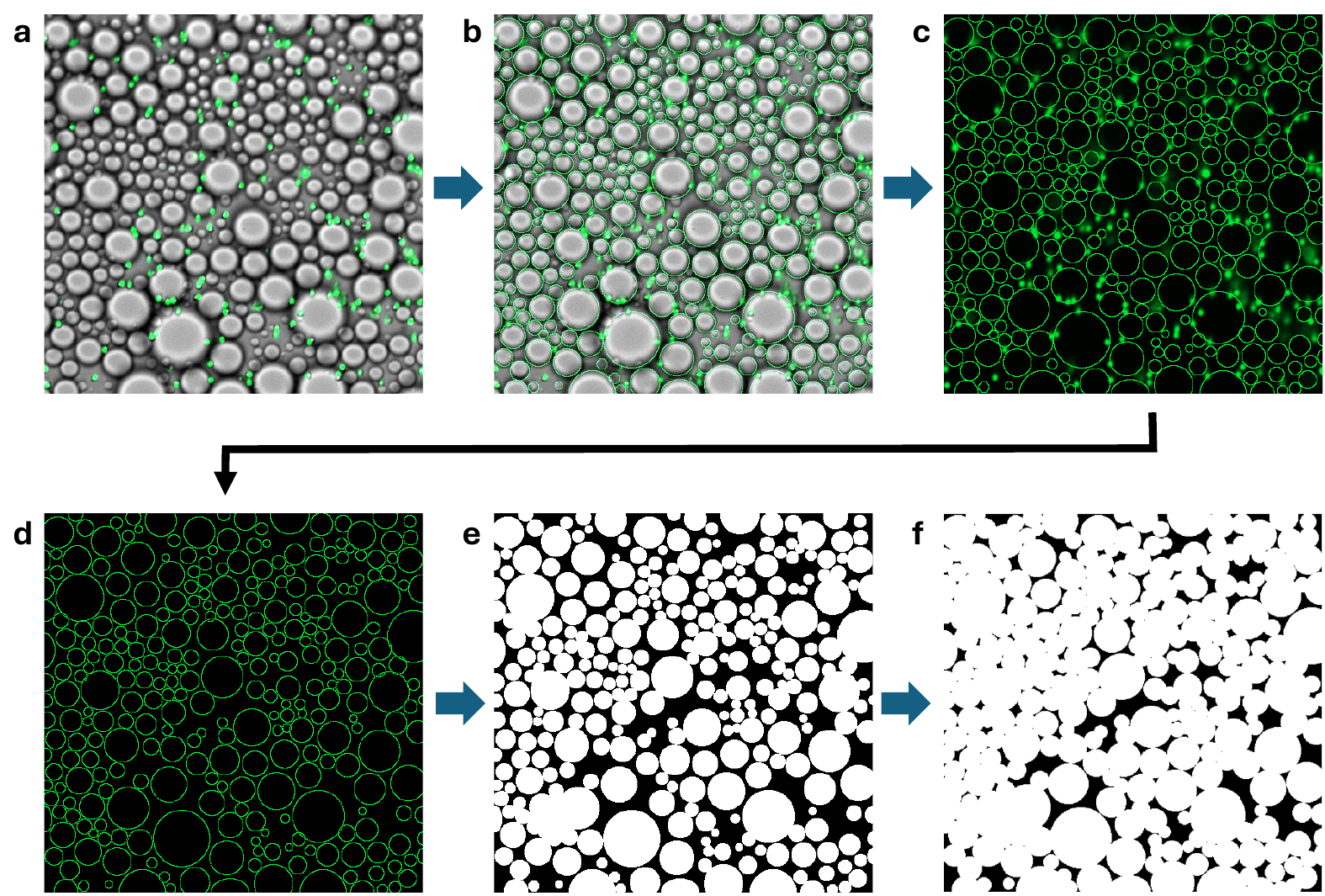}
    \caption{\textbf{Image-processing procedure for extracting the experimental pore size distribution.} (a) Merged BF/GFP image, (b) matrix particles outlined with circles, (c) GFP channel after channel splitting, (d) obtained matrix-particle outlines, (e) binary mask of the pore spaces, and (f) tracer-effective pore spaces.}
\label{fig:exp_pore_measure}
\end{figure}
The experimental pore size distribution was extracted from two-channel microscopy images consisting of bright-field and GFP channels. The bright-field channel was used to visualize the hydrogel matrix particles, whereas the GFP channel contained the fluorescent tracer particles. As shown in Fig.~\ref{fig:exp_pore_measure}a, the two channels were first merged and opened in Fiji so that the matrix particles could be identified together with the tracers. Circular outlines were then manually drawn around the matrix particles on the merged image, as shown in Fig.~\ref{fig:exp_pore_measure}b.

After the matrix particles were outlined, the merged image was split back into individual channels, and only the GFP channel was retained. This channel contained both the fluorescent tracer signal and the matrix-particle boundaries (Fig.~\ref{fig:exp_pore_measure}c). To isolate the boundaries of the matrix particle, the original GFP image containing only the tracer fluorescence was subtracted from the outlined GFP image (Fig.~\ref{fig:exp_pore_measure}d). The interiors of these circles were then filled using the \textit{Fill Holes} function in Fiji, producing a binary mask of the matrix-particle phase (Fig.~\ref{fig:exp_pore_measure}e).

The pore space obtained from the binary mask does not directly represent the pore space experienced by the tracer particles. Since the tracers have a finite radius, their centers cannot access regions within one tracer radius of the matrix-particle boundaries. Therefore, the geometric pore mask was further processed in MATLAB to account for the tracer size. The binary image was first thresholded, and the matrix mask was cleaned by filling holes and removing small artifacts. The geometric pore space was then defined as the complementary region of the matrix mask. A Euclidean distance transform was used to calculate the distance from each pore pixel to the nearest matrix-particle boundary. Only pore pixels farther than one tracer radius from the matrix phase were retained. The resulting mask was defined as the tracer-effective pore space accessible to the tracer center (Fig.~\ref{fig:exp_pore_measure}f). Individual effective pores were identified as connected components of this tracer-effective pore mask.

\subsection{Pore-size characterization in simulation}
In simulations, we characterize the pore space as the regions of the domain not occupied by matrix disks to quantify the geometric constraints imposed by the matrix particles.
We consider a two-dimensional square domain of side length $L = 150~\mu\mathrm{m}$ populated by $N_m$ circular matrix particles with centers $\mathbf{x}_k$ and radii $R_k$. To avoid boundary artifacts, the pore analysis is restricted to a region of interest (ROI) defined as
\begin{equation}
x_1,x_2 \in [r_t, L-r_t],
\end{equation}
where $r_t = 1.0~\mu\mathrm{m}$ is the tracer radius. This restriction ensures that tracer centers remain fully inside the domain during the geometric analysis.

\subsubsection{Spatial discretization}

To resolve the irregular pore geometry generated by the disordered matrix, the ROI is discretized onto a uniform square grid with resolution 
$n_1 = n_2 = 1200$. The resulting grid spacing is
\begin{equation}
\Delta x_1 = \Delta x_2 = \frac{L}{n_x-1} \approx 0.125~\mu\mathrm{m}.
\end{equation}
This spacing corresponds to approximately sixteen grid points across the tracer diameter ($2r_t = 2~\mu\mathrm{m}$), ensuring that pore boundaries and areas are accurately resolved while maintaining computational efficiency. Each grid point represents the center of a square cell with area $\Delta x_1 \Delta x_2$.

\subsubsection{Geometrically accessible space}

A grid point $\mathbf{x}$ is considered geometrically accessible if it lies outside all matrix particles,
\begin{equation}
\|\mathbf{x}-\mathbf{x}_k\| > R_k
\quad \text{for all } k=1,\dots,N_m.
\end{equation}
The resulting binary field defines the free-space geometry within the ROI.

\subsubsection{Local geometric clearance}

To characterize the local size of the pore space, we compute the minimum distance from each accessible point to the nearest matrix boundary,
\begin{equation}
d(\mathbf{x}) =
\min_k \left( \|\mathbf{x}-\mathbf{x}_k\| - R_k \right).
\end{equation}
This distance corresponds to the radius of the largest circle that can be inscribed at position $\mathbf{x}$ without intersecting the matrix particles. Equivalently, this procedure can be interpreted as a morphological dilation of the matrix by a probe radius, a standard method for characterizing disordered porous media.

\begin{figure}
\centering
\includegraphics[width=0.75\textwidth]{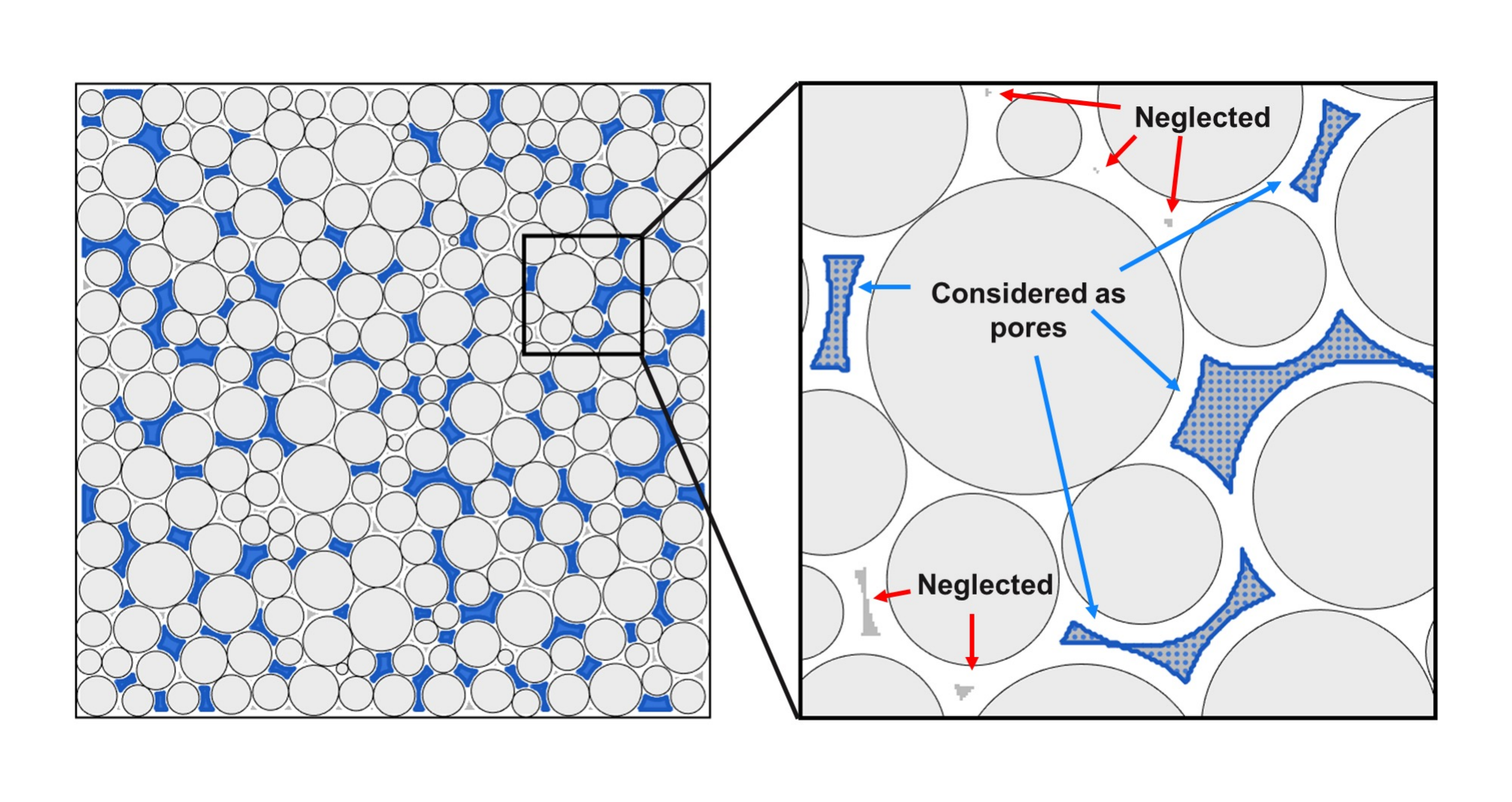}
   \caption{\textbf{Identification of individual pores.}
   Schematic illustration of pore identification in the matrix configuration.}
\label{fig:pore_detect}
\end{figure}

\subsubsection{Identification of individual pores}

Pores are defined as connected regions of geometrically accessible grid points, with connectivity evaluated using nearest-neighbor adjacency (up–down and left–right, excluding diagonal). Each connected component therefore represents a contiguous void formed by the surrounding matrix particles.

The area of each pore $p$ is computed as
\begin{equation}
A_p = N_{\mathrm{grid},p}\,\Delta x_1 \Delta x_2,
\end{equation}
where $N_{\mathrm{grid},p}$ is the number of grid points belonging to pore $p$. This definition naturally captures the irregular shapes of pores as it does not assume any predefined geometry.

To suppress discretization noise and exclude geometrically insignificant voids, pores with areas smaller than $A_{\min} = 3.14~\mu\mathrm{m}^2$ (shown as gray), are excluded from the analysis, as the tracer is too big to fit inside them (Fig.~\ref{fig:pore_detect}).

For interpretability, we additionally report an equivalent pore diameter,
\begin{equation}
d_{\mathrm{eq}} = 2\sqrt{\frac{A_p}{\pi}},
\end{equation}
representing the diameter of a circle with the same area. While several geometric metrics can be defined, pore area provides the most robust descriptor of confinement because it captures the total navigable space rather than a single local length scale.

\subsubsection{Tracer-accessible pore space}

Because tracers have a finite radius, not all geometrically available regions are physically accessible to tracer centers. A point is therefore considered accessible to a tracer center only if
\begin{equation}
d(\mathbf{x}) \ge r_t.
\end{equation}
Equivalently, this condition can be interpreted as inflating each matrix particle by the tracer radius,
\begin{equation}
\|\mathbf{x}-\mathbf{x}_k\| > R_k + r_t.
\end{equation}
Connected regions satisfying this constraint define the tracer-passable pores. For these pores, we report the effective clearance
\begin{equation}
d_{\max}^{(\mathrm{eff})} = d_{\max} - r_t,
\end{equation}
which represents the additional space available beyond the tracer size.

\begin{figure}
\centering
\includegraphics[width=\textwidth]{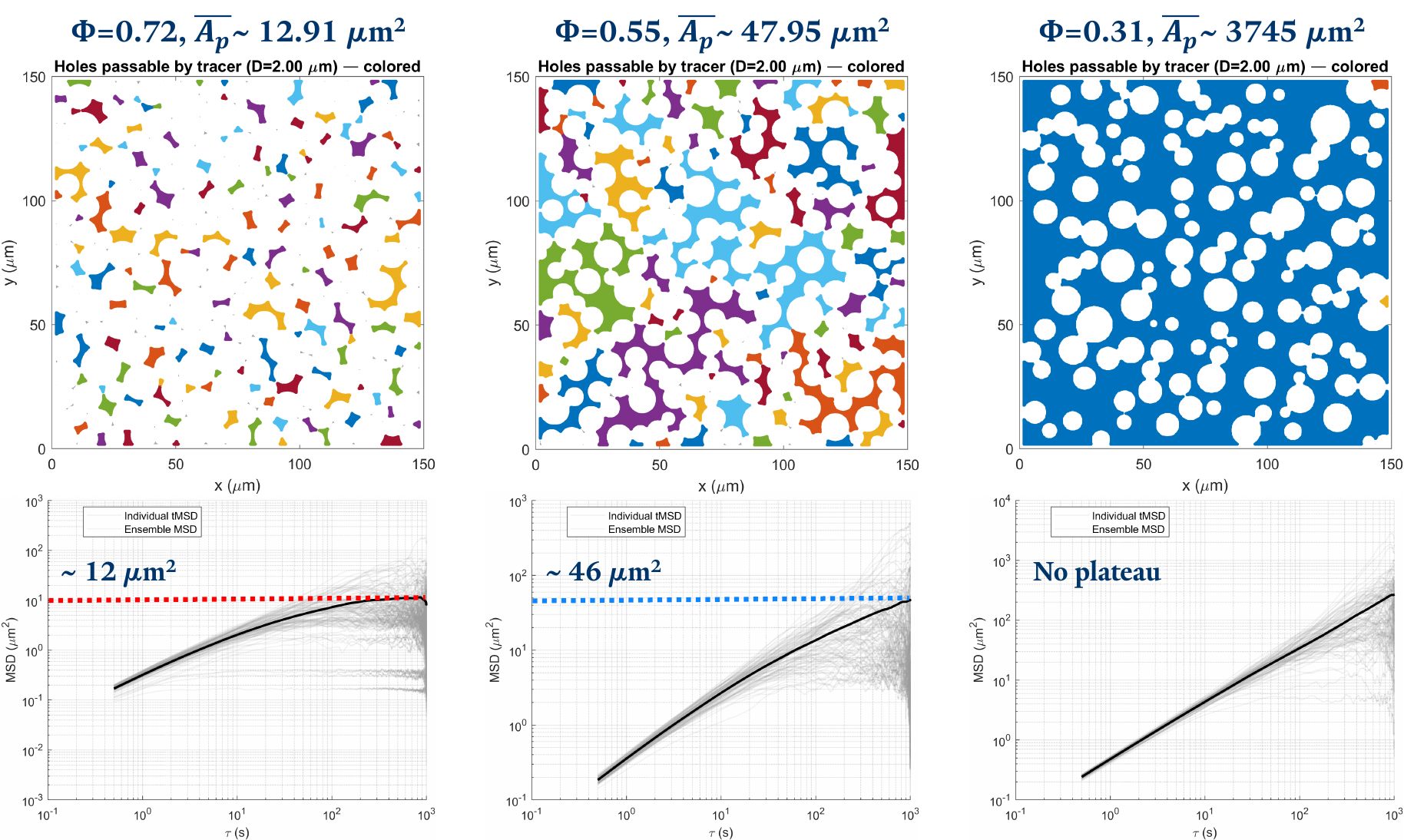}
   \caption{\textbf{Relationship between pore size and tracer dynamics.} Top row: Pores detected for three matrix particle area fractions. Gray indicates pores excluded from the analysis. Bottom row: Gray curves show MSDs from individual tracer trajectories, and the black solid line shows the ensemble-averaged MSD. The dashed line indicates the average pore size.}
\label{fig:pore_size_plateau}
\end{figure}

The resulting pore sizes are averaged over different matrix particle area fractions, and we find that when a plateau exists, it is comparable to the average pore size (Fig.~\ref{fig:pore_size_plateau}).

\begin{figure}
\centering
\includegraphics[width=0.6\textwidth]{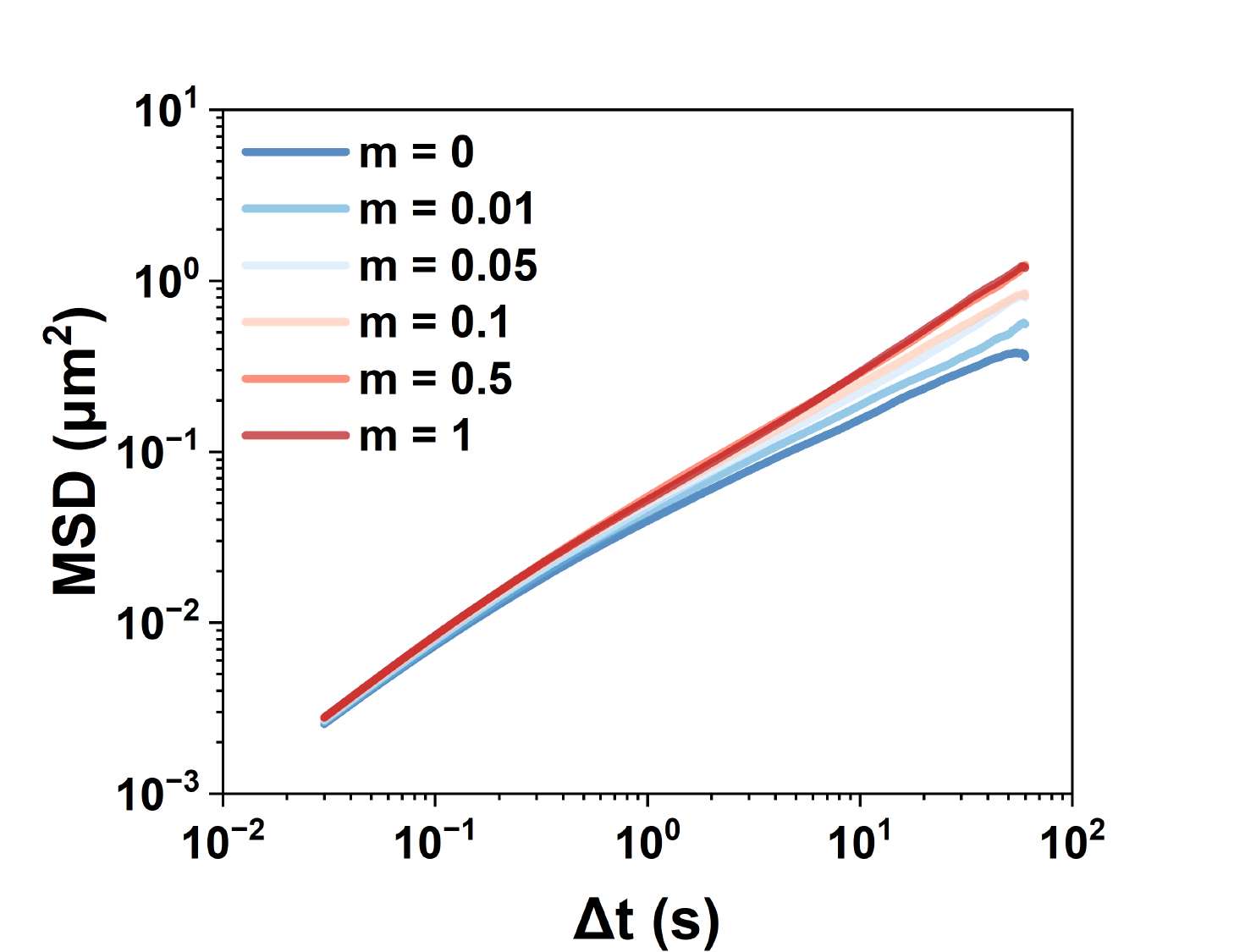}
   \caption{Effect of matrix particle mobility $m$. $m = 0$ means a fixed matrix particle. $m = 1$ means that the matrix particles experience the same local viscosity as the tracer. Changing $m$ most significantly impacts the behavior of the tracer MSD at long $\Delta t$.}
\label{fig:effect_of_mobility}
\end{figure}

\clearpage

\section*{Supplementary movies} 
\paragraph{Caption for Movie S1.}
\textbf{Representative tracer dynamics in a densely packed soft-particle matrix.}
The movie shows the motion of $2~\mu\mathrm{m}$ tracer particles (green) embedded in a suspension of PEGDA hydrogel particles with diameter $\approx 9~\mu\mathrm{m}$. The matrix particle area fraction is $\Phi = 0.72$. The movie is played at $26\times$ real time.

\paragraph{Caption for Movie S2.}
\textbf{Evolution of particle packing in the simulation.}
The movie shows the packing of matrix particles with geometric parameters $\bar{R} = 4.56$, $p = 0.286$, and $\Phi = 0.717$.

\paragraph{Caption for Movie S3.}
\textbf{Simulated motion of tracer particles in a densely packed soft-particle matrix.}
The movie shows the trajectories of 200 tracer particles in a simulated matrix with geometric parameters $\bar{R} = 4.56$, $p = 0.286$, and $\Phi = 0.717$.


\bibliographystyle{unsrt}
\bibliography{ref}

@article{schnyder2018crowding,
  title={Crowding of interacting fluid particles in porous media through molecular dynamics: Breakdown of universality for soft interactions},
  author={Schnyder, Simon K and Horbach, J{\"u}rgen},
  journal={Phys. Rev. Lett.},
  volume={120},
  number={7},
  pages={078001},
  year={2018},
  publisher={APS}
}

@article{wang2021time,
  title={Time averaging and emerging nonergodicity upon resetting of fractional Brownian motion and heterogeneous diffusion processes},
  author={Wang, Wei and Cherstvy, Andrey G and Kantz, Holger and Metzler, Ralf and Sokolov, Igor M},
  journal={Phys. Rev. E},
  volume={104},
  number={2},
  pages={024105},
  year={2021},
  publisher={APS}
}

@article{gu2024ab,
  title={Ab initio uncertainty quantification in scattering analysis of microscopy},
  author={Gu, Mengyang and He, Yue and Liu, Xubo and Luo, Yimin},
  journal={Phys. Rev. E},
  volume={110},
  number={3},
  pages={034601},
  year={2024},
  publisher={APS}
}

@article{sarfati2021enhanced,
  title={Enhanced diffusive transport in fluctuating porous media},
  author={Sarfati, Rapha{\"e}l and Calderon, Christopher P and Schwartz, Daniel K},
  journal={ACS nano},
  volume={15},
  number={4},
  pages={7392--7398},
  year={2021},
  publisher={ACS Publications}
}

@article{kuimova2009imaging,
  title={Imaging intracellular viscosity of a single cell during photoinduced cell death},
  author={Kuimova, Marina K and Botchway, Stanley W and Parker, Anthony W and Balaz, Milan and Collins, Hazel A and Anderson, Harry L and Suhling, Klaus and Ogilby, Peter R},
  journal={Nature chemistry},
  volume={1},
  number={1},
  pages={69--73},
  year={2009},
  publisher={Nature Publishing Group UK London}
}

@article{betterton2022new,
  title={A new view of how cytoplasmic viscosity affects microtubule dynamics},
  author={Betterton, Meredith D},
  journal={Developmental Cell},
  volume={57},
  number={4},
  pages={419--420},
  year={2022},
  publisher={Elsevier}
}

@article{berret2016local,
  title={Local viscoelasticity of living cells measured by rotational magnetic spectroscopy},
  author={Berret, J-F},
  journal={Nature communications},
  volume={7},
  number={1},
  pages={10134},
  year={2016},
  publisher={Nature Publishing Group UK London}
}

@article{luo2025optimizing,
  title={Optimizing gelation time for cell shape control through active learning},
  author={Luo, Yuxin and Chen, Juan and Gu, Mengyang and Luo, Yimin},
  journal={Soft Matter},
  volume={21},
  number={5},
  pages={970--981},
  year={2025},
  publisher={Royal Society of Chemistry}
}

@article{xue2022hopping,
  title={Hopping behavior mediates the anomalous confined diffusion of nanoparticles in porous hydrogels},
  author={Xue, Chundong and Huang, Yirong and Zheng, Xu and Hu, Guoqing},
  journal={The Journal of Physical Chemistry Letters},
  volume={13},
  number={45},
  pages={10612--10620},
  year={2022},
  publisher={ACS Publications}
}

@article{hale2009resolving,
  title={Resolving the role of actoymyosin contractility in cell microrheology},
  author={Hale, Christopher M and Sun, Sean X and Wirtz, Denis},
  journal={PloS one},
  volume={4},
  number={9},
  pages={e7054},
  year={2009},
  publisher={Public Library of Science San Francisco, USA}
}

@article{weihs2006bio,
  title={Bio-microrheology: a frontier in microrheology},
  author={Weihs, Daphne and Mason, Thomas G and Teitell, Michael A},
  journal={Biophys. J.},
  volume={91},
  number={11},
  pages={4296--4305},
  year={2006},
  publisher={Elsevier}
}

@article{wirtz2009particle,
  title={Particle-tracking microrheology of living cells: principles and applications},
  author={Wirtz, Denis},
  journal={Annual Review of Biophysics},
  volume={38},
  number={1},
  pages={301--326},
  year={2009},
  publisher={Annual Reviews}
}

@article{burakov2021persistent,
  title={Persistent growth of microtubules at low density},
  author={Burakov, Anton and Vorobjev, Ivan and Semenova, Irina and Cowan, Ann and Carson, John and Wu, Yi and Rodionov, Vladimir},
  journal={Mol. Biol. Cell.},
  volume={32},
  number={5},
  pages={435--445},
  year={2021},
  publisher={The American Society for Cell Biology}
}

@article{berry2014anomalous,
  title={Anomalous diffusion due to hindering by mobile obstacles undergoing Brownian motion or Orstein-Ulhenbeck processes},
  author={Berry, Hugues and Chat{\'e}, Hugues},
  journal={Phys. Rev. E},
  volume={89},
  number={2},
  pages={022708},
  year={2014},
  publisher={APS}
}

@article{cho2012effect,
  title={Effect of polydispersity on diffusion in random obstacle matrices},
  author={Cho, Hyun Woo and Kwon, Gyemin and Sung, Bong June and Yethiraj, Arun},
  journal={Phys. Rev. Lett.},
  volume={109},
  number={15},
  pages={155901},
  year={2012},
  publisher={APS}
}

@article{zhou2008macromolecular,
  title={Macromolecular crowding and confinement: biochemical, biophysical, and potential physiological consequences},
  author={Zhou, Huan-Xiang and Rivas, Germ{\'a}n and Minton, Allen P},
  journal={Annu. Rev. Biophys.},
  volume={37},
  number={1},
  pages={375--397},
  year={2008},
  publisher={Annual Reviews}
}

@article{hofling2013anomalous,
  title={Anomalous transport in the crowded world of biological cells},
  author={H{\"o}fling, Felix and Franosch, Thomas},
  journal={Reports on Progress in Physics},
  volume={76},
  number={4},
  pages={046602},
  year={2013},
  publisher={IOP Publishing}
}

@article{skinner2013localization,
  title={Localization dynamics of fluids in random confinement},
  author={Skinner, Thomas OE and Schnyder, Simon K and Aarts, Dirk GAL and Horbach, J{\"u}rgen and Dullens, Roel PA},
  journal={Physical Review Letters},
  volume={111},
  number={12},
  pages={128301},
  year={2013},
  publisher={APS}
}

@article{kurzidim2009single,
  title={Single-particle and collective slow dynamics of colloids in porous confinement},
  author={Kurzidim, Jan and Coslovich, Daniele and Kahl, Gerhard},
  journal={Physical review letters},
  volume={103},
  number={13},
  pages={138303},
  year={2009},
  publisher={APS}
}

@article{schnyder2017dynamic,
  title={Dynamic heterogeneities and non-Gaussian behavior in two-dimensional randomly confined colloidal fluids},
  author={Schnyder, Simon K and Skinner, Thomas OE and Thorneywork, Alice L and Aarts, Dirk GAL and Horbach, J{\"u}rgen and Dullens, Roel PA},
  journal={Phys. Rev. E},
  volume={95},
  number={3},
  pages={032602},
  year={2017},
  publisher={APS}
}

@article{koppel1972analysis,
  title={Analysis of macromolecular polydispersity in intensity correlation spectroscopy: the method of cumulants},
  author={Koppel, Dennis E},
  journal={The Journal of Chemical Physics},
  volume={57},
  number={11},
  pages={4814--4820},
  year={1972},
  publisher={American Institute of Physics}
}

@article{sobol1990sensitivity,
  title={On sensitivity estimation for nonlinear mathematical models},
  author={Sobol', Ilya M },
  journal={Matematicheskoe Modelirovanie},
  volume={2},
  number={1},
  pages={112--118},
  year={1990},
  publisher={Russian Academy of Sciences, Branch of Mathematical Sciences}
}

@article{vicsek1995novel,
  title={Novel type of phase transition in a system of self-driven particles},
  author={Vicsek, Tam{\'a}s and Czir{\'o}k, Andr{\'a}s and Ben-Jacob, Eshel and Cohen, Inon and Shochet, Ofer},
  journal={Phys. Rev. Lett.},
  volume={75},
  number={6},
  pages={1226},
  year={1995},
  publisher={APS}
}

@article{crocker1996methods,
  title={Methods of digital video microscopy for colloidal studies},
  author={Crocker, John C and Grier, David G},
  journal={J. Colloid Interface Sci.},
  volume={179},
  number={1},
  pages={298--310},
  year={1996},
  publisher={Elsevier}
}

@article{lippincott2003photobleaching,
  title={Photobleaching and photoactivation: following protein dynamics in living cells.},
  author={Lippincott-Schwartz, Jennifer and Altan-Bonnet, Nihal and Patterson, George H},
  journal={Nature Cell Biology},
  pages={S7--14},
  year={2003}
}

@article{pastore2021rapid,
  title={Rapid Fickian yet non-Gaussian diffusion after subdiffusion},
  author={Pastore, Raffaele and Ciarlo, Antonio and Pesce, Giuseppe and Greco, Francesco and Sasso, Antonio},
  journal={Phys. Rev. Lett.},
  volume={126},
  number={15},
  pages={158003},
  year={2021},
  publisher={APS}
}

@article{pujol2017sensitivity,
  title={Sensitivity: global sensitivity analysis of model outputs},
  author={Pujol, G and Iooss, B and Janon, A and Boumhaout, K and Da Veiga, S and Fruth, J and Gilquin, L and Guillaume, J and Le Gratiet, L and Lemaitre, P and others},
  journal={R package version},
  volume={1},
  number={0},
  pages={--},
  year={2017}
}

@article{collins2019nonuniform,
  title={Nonuniform crowding enhances transport},
  author={Collins, Matthew and Mohajerani, Farzad and Ghosh, Subhadip and Guha, Rajarshi and Lee, Tae-Hee and Butler, Peter J and Sen, Ayusman and Velegol, Darrell},
  journal={ACS nano},
  volume={13},
  number={8},
  pages={8946--8956},
  year={2019},
  publisher={ACS Publications}
}

@article{reits2001fixed,
  title={From fixed to FRAP: measuring protein mobility and activity in living cells},
  author={Reits, Eric AJ and Neefjes, Jacques J},
  journal={Nature cell biology},
  volume={3},
  number={6},
  pages={E145},
  year={2001},
  publisher={Nature Publishing Group}
}

@article{brenner1961slow,
  title={The slow motion of a sphere through a viscous fluid towards a plane surface},
  author={Brenner, Howard},
  journal={Chemical engineering science},
  volume={16},
  number={3-4},
  pages={242--251},
  year={1961},
  publisher={Elsevier}
}

@article{ning2021diffusion,
  title={Diffusion of colloidal particles in model porous media},
  author={Ning, Luhui and Liu, Peng and Ye, Fangfu and Yang, Mingcheng and Chen, Ke},
  journal={Phys. Rev. E},
  volume={103},
  number={2},
  pages={022608},
  year={2021},
  publisher={APS}
}

@article{goldman1967slow,
  title={Slow viscous motion of a sphere parallel to a plane wall—I Motion through a quiescent fluid},
  author={Goldman, Arthur Joseph and Cox, Raymond G and Brenner, Howard},
  journal={Chemical engineering science},
  volume={22},
  number={4},
  pages={637--651},
  year={1967},
  publisher={Elsevier}
}

@article{mason2000estimating,
  title={Estimating the viscoelastic moduli of complex fluids using the generalized {S}tokes--{E}instein equation},
  author={Mason, Thomas G},
  journal={Rheol. Acta},
  volume={39},
  number={4},
  pages={371--378},
  year={2000},
  publisher={Springer}
}

@book{rasmussen2006gaussian,
  title={Gaussian processes for machine learning},
  author={Rasmussen, Carl Edward},
  year={2006},
  publisher={MIT Press}
}

@article{guan2014even,
  title={Even hard-sphere colloidal suspensions display Fickian yet non-Gaussian diffusion},
  author={Guan, Juan and Wang, Bo and Granick, Steve},
  journal={ACS nano},
  volume={8},
  number={4},
  pages={3331--3336},
  year={2014},
  publisher={ACS Publications}
}

@article{skaug2015hindered,
  title={Hindered nanoparticle diffusion and void accessibility in a three-dimensional porous medium},
  author={Skaug, Michael J and Wang, Liang and Ding, Yifu and Schwartz, Daniel K},
  journal={ACS nano},
  volume={9},
  number={2},
  pages={2148--2156},
  year={2015},
  publisher={ACS Publications}
}

@article{chate2008modeling,
  title={Modeling collective motion: variations on the Vicsek model},
  author={Chat{\'e}, Hugues and Ginelli, Francesco and Gr{\'e}goire, Guillaume and Peruani, Fernando and Raynaud, Franck},
  journal={The European Physical Journal B},
  volume={64},
  number={3},
  pages={451--456},
  year={2008},
  publisher={Springer}
}

@article{gao2009accurate,
  title={Accurate detection and complete tracking of large populations of features in three dimensions},
  author={Gao, Yongxiang and Kilfoil, Maria L},
  journal={Opt. Exp.},
  volume={17},
  number={6},
  pages={4685--4704},
  year={2009},
  publisher={Optical Society of America}
}

@article{sabri2020elucidating,
  title={Elucidating the origin of heterogeneous anomalous diffusion in the cytoplasm of mammalian cells},
  author={Sabri, Adal and Xu, Xinran and Krapf, Diego and Weiss, Matthias},
  journal={Physical Review Letters},
  volume={125},
  number={5},
  pages={058101},
  year={2020},
  publisher={APS}
}

@article{weiss2013single,
  title={Single-particle tracking data reveal anticorrelated fractional Brownian motion in crowded fluids},
  author={Weiss, Matthias},
  journal={Physical Review E—Statistical, Nonlinear, and Soft Matter Physics},
  volume={88},
  number={1},
  pages={010101},
  year={2013},
  publisher={APS}
}

@article{ernst2014probing,
  title={Probing the type of anomalous diffusion with single-particle tracking},
  author={Ernst, Dominique and K{\"o}hler, J{\"u}rgen and Weiss, Matthias},
  journal={Physical Chemistry Chemical Physics},
  volume={16},
  number={17},
  pages={7686--7691},
  year={2014},
  publisher={Royal Society of Chemistry}
}

@article{destrian2026cytoplasmic,
  title={Cytoplasmic crowding acts as a porous medium reducing macromolecule diffusion},
  author={Destrian, Olivier and Moisan, Nicolas and M{\`e}ge, Ren{\'e}-Marc and Ladoux, Benoit and Goyeau, Benoit and Chabanon, Morgan},
  journal={Proceedings of the National Academy of Sciences},
  volume={123},
  number={4},
  pages={e2519599123},
  year={2026},
  publisher={National Academy of Sciences}
}

@article{ermak1978brownian,
  title={Brownian dynamics with hydrodynamic interactions},
  author={Ermak, Donald L and McCammon, J Andrew},
  journal={The Journal of chemical physics},
  volume={69},
  number={4},
  pages={1352--1360},
  year={1978},
  publisher={AIP Publishing}
}

@article{zeitz2017active,
  title={Active Brownian particles moving in a random Lorentz gas},
  author={Zeitz, Maria and Wolff, Katrin and Stark, Holger},
  journal={The European Physical Journal E},
  volume={40},
  number={2},
  pages={23},
  year={2017},
  publisher={Springer}
}

@article{wong2004anomalous,
  title={Anomalous diffusion probes microstructure dynamics of entangled F-actin networks},
  author={Wong, Ian Y and Gardel, Margaret L and Reichman, David R and Weeks, Eric R and Valentine, Megan T and Bausch, Andreas R and Weitz, David A},
  journal={Physical review letters},
  volume={92},
  number={17},
  pages={178101},
  year={2004},
  publisher={APS}
}

@article{bronstein2009transient,
  title={Transient anomalous diffusion of telomeres in the nucleus of mammalian cells},
  author={Bronstein, Irena and Israel, Yonatan and Kepten, Eldad and Mai, Sabine and Shav-Tal, Yaron and Barkai, E and Garini, Yuval},
  journal={Physical review letters},
  volume={103},
  number={1},
  pages={018102},
  year={2009},
  publisher={APS}
}

@article{golding2006physical,
  title={Physical nature of bacterial cytoplasm},
  author={Golding, Ido and Cox, Edward C},
  journal={Physical review letters},
  volume={96},
  number={9},
  pages={098102},
  year={2006},
  publisher={APS}
}

@article{tolic2004anomalous,
  title={Anomalous diffusion in living yeast cells},
  author={Toli{\'c}-N{\o}rrelykke, Iva Marija and Munteanu, Emilia-Laura and Thon, Genevieve and Oddershede, Lene and Berg-S{\o}rensen, Kirstine},
  journal={Physical review letters},
  volume={93},
  number={7},
  pages={078102},
  year={2004},
  publisher={APS}
}

@article{banks2005anomalous,
  title={Anomalous diffusion of proteins due to molecular crowding},
  author={Banks, Daniel S and Fradin, C{\'e}cile},
  journal={Biophysical journal},
  volume={89},
  number={5},
  pages={2960--2971},
  year={2005},
  publisher={Elsevier}
}

@article{weiss2004anomalous,
  title={Anomalous subdiffusion is a measure for cytoplasmic crowding in living cells},
  author={Weiss, Matthias and Elsner, Markus and Kartberg, Fredrik and Nilsson, Tommy},
  journal={Biophysical journal},
  volume={87},
  number={5},
  pages={3518--3524},
  year={2004},
  publisher={Elsevier}
}

@article{lin2026model,
  title={Model-free estimation in scattering analysis of microscopy},
  author={Lin, Tong and Lee, Jinseok and Helgeson, Matt and Valentine, Megan T. and Luo, Yimin and  Gu, Mengyang},
  journal={arXiv preprint 	arXiv:2605.29424},
  year={2026}
}

@article{bouchaud1990anomalous,
  title={Anomalous diffusion in disordered media: statistical mechanisms, models and physical applications},
  author={Bouchaud, Jean-Philippe and Georges, Antoine},
  journal={Physics reports},
  volume={195},
  number={4-5},
  pages={127--293},
  year={1990},
  publisher={Elsevier}
}

@article{moeendarbary2013cytoplasm,
  title={The cytoplasm of living cells behaves as a poroelastic material},
  author={Moeendarbary, Emad and Valon, L{\'e}o and Fritzsche, Marco and Harris, Andrew R and Moulding, Dale A and Thrasher, Adrian J and Stride, Eleanor and Mahadevan, L and Charras, Guillaume T},
  journal={Nature materials},
  volume={12},
  number={3},
  pages={253--261},
  year={2013},
  publisher={Nature Publishing Group UK London}
}

@article{regner2013anomalous,
  title={Anomalous diffusion of single particles in cytoplasm},
  author={Regner, Benjamin M and Vu{\v{c}}ini{\'c}, Dejan and Domnisoru, Cristina and Bartol, Thomas M and Hetzer, Martin W and Tartakovsky, Daniel M and Sejnowski, Terrence J},
  journal={Biophysical journal},
  volume={104},
  number={8},
  pages={1652--1660},
  year={2013},
  publisher={Elsevier}
}

@article{he2016dynamic,
  title={Dynamic heterogeneity and non-{G}aussian statistics for acetylcholine receptors on live cell membrane},
  author={He, Wei and Song, Hao and Su, Yun and Geng, Ling and Ackerson, Bruce J and Peng, HB and Tong, Penger},
  journal={Nat. Comm.},
  volume={7},
  number={1},
  pages={11701},
  year={2016},
  publisher={Nature Publishing Group UK London}
}

@article{gu2016parallel,
  title={Parallel partial {G}aussian process emulation for computer models with massive output},
  author={Gu, Mengyang and Berger, James O},
  journal={The Annals of Applied Statistics},
  volume={10},
  number={3},
  pages={1317--1347},
  year={2016},
  publisher={Institute of Mathematical Statistics}
}

@article{sentjabrskaja2016anomalous,
  title={Anomalous dynamics of intruders in a crowded environment of mobile obstacles},
  author={Sentjabrskaja, Tatjana and Zaccarelli, Emanuela and De Michele, Cristiano and Sciortino, Francesco and Tartaglia, Piero and Voigtmann, Thomas and Egelhaaf, Stefan U and Laurati, Marco},
  journal={Nat. Comm.},
  volume={7},
  number={1},
  pages={1--8},
  year={2016},
  publisher={Nature Publishing Group}
}

@book{furst2017microrheology,
  title={Microrheology},
  author={Furst, Eric M and Squires, Todd M},
  year={2017},
  publisher={Oxford University Press}
}

@article{bayles2017probe,
  title={Probe microrheology without particle tracking by differential dynamic microscopy},
  author={Bayles, Alexandra V and Squires, Todd M and Helgeson, Matthew E},
  journal={Rheol. Acta},
  volume={56},
  number={11},
  pages={863--869},
  year={2017},
  publisher={Springer}
}

@article{gu2018robustgasp,
  author = {Mengyang Gu and Jesus Palomo and James O. Berger},
  title = {{RobustGaSP: Robust Gaussian Stochastic Process Emulation in
          R}},
  year = {2019},
  journal = {{The R Journal}},
  doi = {10.32614/RJ-2019-011},
  pages = {112--136},
  volume = {11},
  number = {1}
}

@article{gu2018jointly,
  title={Jointly Robust Prior for {G}aussian Stochastic Process in Emulation, Calibration and Variable Selection},
  author={Gu, Mengyang},
  journal={Bayesian Analysis},
  volume={14},
  number={1},
  page={857-885},
  year={2018}
}

@article{gu2021uncertainty,
  title={Uncertainty quantification and estimation in differential dynamic microscopy},
  author={Gu, Mengyang and Luo, Yimin and He, Yue and Helgeson, Matthew E and Valentine, Megan T},
  journal={Phys. Rev.  E},
  volume={104},
  number={3},
  pages={034610},
  year={2021},
  publisher={APS}
}

@article{luo2022high,
  title={High-throughput microscopy to determine morphology, microrheology, and phase boundaries applied to phase separating coacervates},
  author={Luo, Yimin and Gu, Mengyang and Edwards, Chelsea ER and Valentine, Megan T and Helgeson, Matthew E},
  journal={Soft Matter},
  volume={18},
  number={15},
  pages={3063--3075},
  year={2022},
  publisher={Royal Society of Chemistry}
}

@article{wang2012brownian,
  title={When {B}rownian diffusion is not {G}aussian},
  author={Wang, Bo and Kuo, James and Bae, Sung Chul and Granick, Steve},
  journal={Nat. Mater.},
  volume={11},
  number={6},
  pages={481--485},
  year={2012},
  publisher={Nature Publishing Group}
}

@article{leptos2009dynamics,
  title={Dynamics of enhanced tracer diffusion in suspensions of swimming eukaryotic microorganisms},
  author={Leptos, Kyriacos C and Guasto, Jeffrey S and Gollub, Jerry P and Pesci, Adriana I and Goldstein, Raymond E},
  journal={Phys. Rev. Lett.},
  volume={103},
  number={19},
  pages={198103},
  year={2009},
  publisher={APS}
}

@article{wang2009anomalous,
  title={Anomalous yet {B}rownian},
  author={Wang, Bo and Anthony, Stephen M and Bae, Sung Chul and Granick, Steve},
  journal={Proc. Natl. Acad. Sci. U.S.A.},
  volume={106},
  number={36},
  pages={15160--15164},
  year={2009},
  publisher={National Acad Sciences}
}

@article{o2003jamming,
  title={Jamming at zero temperature and zero applied stress: The epitome of disorder},
  author={O’hern, Corey S and Silbert, Leonardo E and Liu, Andrea J and Nagel, Sidney R},
  journal={Phys. Rev. E},
  volume={68},
  number={1},
  pages={011306},
  year={2003},
  publisher={APS}
}

@article{saldanha2025competing,
  title={Competing Crosstalk between Cytoskeletal Filaments Dictates Structure and Superdiffusivity of Microtubules in Live Cells},
  author={Saldanha, Renita and Lan, Yiling and Hehnly, Heidi and McGorty, Ryan J and Robertson-Anderson, Rae M and Patteson, Alison},
  journal={PRX Life},
  volume={3},
  number={3},
  pages={033009},
  year={2025},
  publisher={APS}
}

@article{svitkina2020actin,
  title={Actin cell cortex: structure and molecular organization},
  author={Svitkina, Tatyana M},
  journal={Trends in cell biology},
  volume={30},
  number={7},
  pages={556--565},
  year={2020},
  publisher={Elsevier}
}

@article{ellis2003join,
  title={Join the crowd},
  author={Ellis, R John and Minton, Allen P},
  journal={Nature},
  volume={425},
  number={6953},
  pages={27--28},
  year={2003},
  publisher={Nature Publishing Group UK London}
}

@article{bi2016motility,
  title={Motility-driven glass and jamming transitions in biological tissues},
  author={Bi, Dapeng and Yang, Xingbo and Marchetti, M Cristina and Manning, M Lisa},
  journal={Physical Review X},
  volume={6},
  number={2},
  pages={021011},
  year={2016},
  publisher={APS}
}

@article{mandelbrot1968fractional,
  title={Fractional {B}rownian motions, fractional noises and applications},
  author={Mandelbrot, Benoit B and Van Ness, John W},
  journal={SIAM review},
  volume={10},
  number={4},
  pages={422--437},
  year={1968},
  publisher={SIAM}
}

@article{Bevan2000,
  author  = {Bevan, Michael A. and Prieve, Dennis C.},
  title   = {Hindered diffusion of colloidal particles very near to a wall},
  journal = {The Journal of Chemical Physics},
  year    = {2000},
  volume  = {113},
  number  = {3},
  pages   = {1228--1236},
  doi     = {10.1063/1.481900}
}

@book{KimKarrila1991,
  author    = {Kim, Sangtae and Karrila, Seppo J.},
  title     = {Microhydrodynamics: Principles and Selected Applications},
  year      = {1991},
  publisher = {Butterworth-Heinemann},
  address   = {Boston}
}

\end{document}